# Algorithmic Problem Complexity

**Mark Burgin**

Department of Computer Science
University of California, Los Angeles
405 Hilgard Ave.
Los Angeles, CA 90095

> Universes of virtually unlimited
> complexity can be created in the
> form of computer programs.
> Joseph Weizenbaum

**Abstract:** People solve different problems and know that some of them are simple, some are complex and some insoluble. The main goal of this work is to develop a mathematical theory of algorithmic complexity for problems. This theory is aimed at determination of computer abilities in solving different problems and estimation of resources that computers need to do this. Here we build the part of this theory related to static measures of algorithms. At first, we consider problems for finite words and study algorithmic complexity of such problems, building optimal complexity measures. Then we consider problems for such infinite objects as functions and study algorithmic complexity of these problems, also building optimal complexity measures. In the second part of the work, complexity of algorithmic problems, such as the halting problem for Turing machines, is measured by the classes of automata that are necessary to solve this problem. To classify different problems with respect to their complexity, inductive Turing machines, which extend possibilities of Turing machines, are used. A hierarchy of inductive Turing machines generates an inductive hierarchy of algorithmic problems. Here we specifically consider algorithmic problems related to Turing machines and inductive Turing machines, and find a place for these problems in the inductive hierarchy of algorithmic problems.

## 1. Introduction

One of the scientific reflections of efficiency is complexity. Kolmogorov, or algorithmic, complexity has become an important and popular tool in computer science, programming, probability theory, statistics, and information theory. Algorithmic complexity has found applications in medicine, biology, neurophisiology, physics, economics, hardware and software engineering. In biology, algorithmic complexity is used for estimation of protein identification [24, 25]. In physics, problems of quantum gravity are analyzed based on algorithmic complexity of given object. In particular, the algorithmic complexity of the Schwarzschild black hole is estimated [26, 27]. In [3], algorithmic complexity is applied to chaotic dynamics. In [56, 57], the inclusion of algorithmic complexity and randomness in the definition of physical entropy allows the author to get a formulation of thermodynamics. In [37], Kolmogorov complexity is applied to problems in mechanics. In [53], the author discusses what can be the algorithmic complexity of the whole universe. The main problem with this discussion is that the author identifies physical universe with physical models of this universe. To get valid results on this issue, it is necessary to define algorithmic complexity for physical systems because conventional algorithmic complexity is defined only for such symbolic objects as words and texts [15, 40]. Then it is necessary to show that there is a good correlation between algorithmic complexity of the universe and algorithmic complexity of its model used by the author of [53].

In economics, a new approach to understanding of the complex behavior of financial markets using algorithmic complexity is developed [42]. In neurophysiology, algorithmic complexity is used to measure characteristics of brain functions [50]. Algorithmic complexity has been useful in the development of software metrics and other problems of software engineering [12, 22, 39]. In [17], algorithmic complexity is used to study low-bandwidth denial of service attacks that exploit algorithmic deficiencies in many common applications' data structures.

Thus, we see that algorithmic complexity is a frequent word in present days' scientific literature, in various fields and with diverse meanings, appearing in some contexts as a precise concept of algorithmic complexity, while being a vague idea of complexity in general in other texts. The reason for this is that people study and create more and more complex systems.

Algorithmic complexity in its classical form gives an estimate how many bits of information we need to restore a given text by algorithms from a given class. Conventional algorithmic complexity deals with recursive algorithms, such as Turing machines. Inductive algorithmic complexity involves such superrecursive algorithms as inductive Turing machines. However, there are many other problems that people solve with computers, utilizing algorithms, and there are other resources that are measured not in bits and are essential for computation. Thus, it is useful to consider similar measures of complexity for other problems and other kinds of resources. Such a much more general algorithmic complexity that estimates arbitrary resources necessary to compute some word/text has been introduced and studied in an axiomatic setting in [5, 8, 10]. This generalized algorithmic complexity is built as a measure dual to a static complexity measure of algorithms and computations in the sense of [10]. While direct complexity measures, such as the length of a program, time of computation or space (used memory) of computation, characterize algorithms/machines/programs, dual complexity measures are related to problems solved by these algorithms/machines/programs and to results of their functioning. Duality in this work is restricted to static direct complexity measures, such as the length of an algorithm or program. The classical Kolmogorov/algorithmic complexity of a finite object (usually, a word) is obtained from the generalized algorithmic complexity when the dual static complexity measure of algorithms/programs is the length of the algorithm/program. Dual measures include other versions of Kolmogorov complexity: also uniform complexity, prefix complexity, monotone complexity, process complexity, conditional Kolmogorov complexity, time-bounded Kolmogorov complexity, space-bounded Kolmogorov complexity, conditional resource-bounded Kolmogorov complexity, time-bounded prefix complexity, resource-bounded Kolmogorov complexity, as well as inductive algorithmic complexity, communication complexity, circuit complexity, etc. [1, 10, 13]. All these measures evaluate complexity of the problem of building/computing a finite word.

The main goal of this work is to study algorithmic complexity not only for this problem but also for arbitrary problems. Some cases of general problem complexity have been considered before. For instance, problem complexity in software engineering is analyzed in [12, 14]. The concept of problem complexity is examined in [55]. An interesting case of problem complexity is studied by Kreinovich [36]. Here we develop a general theory of algorithmic problem complexity.

In addition, complexity here (Section 4) is measured not as an absolute property, but is relativized with respect to a class from which algorithms for construction/computation are taken. More powerful algorithms allow one to decrease complexity of computation and construction.

It is necessary to remark that there are different types of problem complexity. Descriptive complexity reflects complexity of problem formulation. Constructive complexity reflects complexity of problem solution. The complexity of a problem description often differs from the complexity of its solution. Simple problems, i.e., problems that have short descriptions, may have only complex solutions, i.e., they demand long proofs or a lot of computations. Moreover, as Juedes and Lutz proved [33], many important problems that have hard solutions (those that are P-complete for ESPACE) have low problem complexity, that is, their Kolmogorov complexity or algorithmic information is rather low.

Davies [21] gives a recent example of such a problem in mathematics. He writes: "*A problem* [classification of finite simple groups] *that can be formulated in a few sentences has a solution more than ten thousand pages long. The proof has never been written down in its entirety, may never be written down, and as presently envisaged would not be comprehensible to any single individual. The result is important, and has been used in a wide variety of other problems in group theory, but it might not be correct.*"

Moreover, there are problems formulated in one sentence that have infinite complexity of solution. For instance, the Halting Problem for Turing machines has the following formulation:

*Find a Turing machine that given a word x and a description of a Turing machine T, gives output* 1 *when T halts, starting with input x, and output* 0 *when T does not halt, starting with input x.*

As this problem is unsolvable, it has infinite algorithmic complexity in the class of all Turing machines or any class of recursive algorithms.

All this shows that there are different hierarchies of problems and it is important to know their complexity.

In Section 2, which goes after Introduction, a classification of problems is developed, separating classes of detection, construction and preservation problems. To achieve sufficient generality, we classify problems without precise formalization because formalized mathematical models can distort the real situation. Some researchers forget that computer science or physics is not a pure mathematics. It is necessary to preserve connections to reality. Otherwise, it is possible to come to such paradoxical results as writing that an accepting Turing machine is a model of a computer. Some even try to prove that it is possible to model any computer by a Turing machine. However, we know that an accepting automaton (acceptor) does not give outputs. At the same time, any normal computer is a transducer, which gives outputs when solves problems.

We study here difference between problems because in some textbooks, it is written that a problem in automata theory is the question of deciding whether a given string is a member of some particular language. This statement tells students that automata theory have very little in common with computers because computers solve many problems from real life that have nothing to do with membership in some language.

The main emphasis in this paper is made on construction problems as problems from two other classes can be reduced to construction problems. After this, we embark on the study of problem complexity. At first, we consider problems for finite words and study (Sections 3 and 4) algorithmic complexity of such problems, building optimal complexity measures. The basic archetype of problem complexity is Kolmogorov complexity. From the beginning, Kolmogorov complexity was developed in the class of all Turing machines as a maximal class of algorithms. The aim of Kolmogorov complexity introduction was to ground probability theory and information theory, creating the new approach based on algorithms. After some experimentation with complexity measures, this goal was achieved. The new theories became very popular, although they did not substitute either the classical probability theory, which was grounded before by Kolmogorov [34] on the base of measure theory, or Shannon's information theory.

It is useful to note that the attempt to define an appropriate concept of randomness was unsuccessful in the setting of the initial Kolmogorov complexity. It turned out that the original definition was not relevant for that goal. To get a correct definition of a random infinite sequence, it was necessary to restrict the class of utilized algorithms. That is why Kolmogorov complexity was defined and studied for various classes of subrecursive algorithms. For example, researchers discussed different reasons for restricting power of the device used for computation when estimating the minimal complexity. This was the first indication that it is necessary to consider algorithmic complexity for different classes of algorithms as it is done, for example, in [5]. Correspondingly, in Section 3, we consider problem complexity in the class of all Turing machines, while in Section 4, we extend this concept for an arbitrary class of algorithms that has universal algorithms.

Then (in Section 5) we consider problems for such infinite objects as functions and study algorithmic complexity of these problems, elaborating optimal complexity measures. Kolmogorov complexity for infinite strings was considered by different authors (cf., for example, [4, 47, 52]). Problem complexity of functions encompasses Kolmogorov complexity for infinite strings as a particular case.

Complexity of algorithmic problems, such as the halting problem for Turing machines, is used (in Sections 7 and 8) to build an inductive hierarchy of problems and to find places in this hierarchy for popular algorithmic problems for Turing machines and other recursive algorithms. Examples of such problems are the Halting Problem, Totality Problem (whether a Turing machine gives a result for all inputs), Emptiness Problem (whether a Turing machine never gives a result), and Infinity Problem (whether a Turing machine gives a result for infinitely many inputs). Levels of this hierarchy are measured by the classes of automata necessary to solve this problem. To classify different problems with respect to their complexity, inductive Turing machines, which extend possibilities of Turing machines, are used. A hierarchy of inductive Turing machines described in [7] generates an inductive hierarchy of algorithmic problems. We find a place in the inductive hierarchy of algorithmic problems for considered in Section 7 algorithmic problems related to Turing machines and inductive Turing machines. Some results from Sections 7 and 8 were published in [11] without proofs.

**Basic denotations and definitions**

If $A$ is a set (alphabet), then $A^*$ is the set of all finite strings (words) of elements from $A$.

For a word $x$ from $A^*$, $l(x)$ is the *length* of (number of symbols in) $x$.

$\varepsilon$ denotes the empty word.

If $M$ is an automaton (e.g., Turing machine, inductive Turing machine, random access machine, etc.) and $x$ is a word in the alphabet of this automaton, then $M(x)$ denotes the result of computation of $M$ with the input $x$ when this result exists and $M(x) = *$ when $M$ gives no result being applied to $x$.

$T$ denotes the set of all Turing machines with a fixed alphabet.

**c**: $T \to A^*$ is a (effective) codification of Turing machines such that it possible to reconstruct any Turing machine $T$ by its code **c**($T$). It is possible to find such codifications, for example, in [10].

$<>$: $(A^*) \times (A^*) \to A^*$ is a pairing function that corresponds to each pair $(w, v)$ of words in the alphabet $A$ the word $<w, v>$ in the same alphabet so that different pairs are mapped into different words. It is possible to find how to build pairing functions, for example, in [10].

P($x$) $_R\!\Rightarrow$ Q($x$) means that P($x$) recursively implies Q($x$), i.e., there is a Turing machine $T$ with an oracle for P($x$) such that $T$ decides Q($x$).

**2. Problems and their complexity**

People build computers, develop networks, and perform computations to solve different problems. Thus, to estimate what problems are solvable with the given means and what resources we need to solve a given problem, measures of problem complexity are used.

Usually the following concept is utilized.

**Definition 2.1**. The *complexity of a problem* is the least amount of resources required for solution of this problem.

However, the word *solution* has different meanings. For instance in [12], three kinds of solutions are considered: final, intermediate, and start solutions. Each of these solutions has two forms: static as the obtained result and dynamic as a process that brings us to this result. Here we do not go into these details.

To construct a measure for problem complexity, we build a goal-oriented classification of problems. According to it, there are three main types of problems: *detection*, *construction*, and *preservation*. Each of them has three subtypes.

**Definition 2.2.** A *detection problem* is aimed at detecting something.

Detection problems have the following subtypes:

- A *decision* or *test problem* is to find whether a given object $x$ satisfies a prescribed condition (has a property) $P(x)$.

- A *selection problem* is to select/choose an object $x$ from a given domain $X$ such that $x$ satisfies a prescribed condition (has a property) $P(x)$.

- A (*specified*) *search problem* is to find an object $x$ (in a specified domain $X$) such that $x$ satisfies a prescribed condition (has a property) $P(x)$.

**Definition 2.3.** A *construction problem* is aimed at building or transforming something, or more exactly, it demands to build an $x$ such that $x$ satisfies $P(x)$.

Construction problems have the following subtypes:

- A *reproduction problem* is to build an object $x$ that satisfies the following condition $P(x)$: $x$ is similar to (the same as) a given object $y$.

- A *production problem* is to build by a given technique (procedure, algorithm) an object $x$ that satisfies a prescribed condition (has a property) $P(x)$.

- An *invention problem* is to build an object *x* that satisfies a prescribed condition (has a property) P(*x*) where P(*x*) gives only some properties of *x* and does not specifies how to build it.

**Definition 2.4.** A *preservation/sustaining* problem is aimed at preserving something (a process, data, knowledge, environment, etc.).

Preservation problems have the following subtypes:

- *Abstinence* means to withdraw all our impact from the object we want to preserve.

- *Support* means to provide conditions for preservation, involving some action.

- *Protection* means to withdraw all impact that can change (damage) the object we want to preserve from this object.

Detection and construction problems form the class of *acquisition problems*.

**Definition 2.5.** An *acquisition problem* consists of three parts: absence (may be potential) of some object, understanding of this absence, and a feeling of a need for this object.

Such absent object may be some information, for example, what weather will be tomorrow, or some physical object such as a house or car.

Let us consider some examples.

**Example 2.1.** Find a Turing machine *T* that tests if a word *w* belongs to a formal language *L*. This is a detection problem, or more exactly, a search problem.

**Example 2.2.** Build an automaton *A* that tests if a word *w* belongs to a formal language *L*. This is a construction problem, or more exactly, an invention problem.

**Example 2.3.** Test if a word *w* belongs to a formal language *L*. This is a detection problem, or more exactly, a decision/test problem.

There are definite relations between different types of problems.

**Proposition 2.1.** It is possible to reduce detection problems to construction problems.

Proof. a) Search problem reduction:

If we have a search problem *Q*, then we can change it to the following construction problem: "Find what you need and then build a copy." Another way of reduction is to build an indication (or membership) function or a partial indication (or partial membership) function. An indication function is equal to one when the object satisfies the conditions of the problem and to zero when an object does not satisfy these conditions. A partial indication function is equal to one when the object satisfies the conditions of the problem and is undefined otherwise. Having an indication function $f(x)$, we can find a necessary $x$ by computing values of $f(x)$. This is a construction problem. The object $a$ for which $f(a) = 1$ gives us a solution to the initial problem.

b) Selection problem reduction:

If we have a selection problem $Q$, then we can change it to the following construction problem: "Select what you need and then build a copy." Another way of reduction is, as in the previous case, to build an indication function or a partial indication function and compute its values. The value 1 will indicate the object we need.

c) Test problem reduction:

If we have a test problem $Q$ that asks to find whether a given object $x$ satisfies a prescribed condition $P(x)$, then we can change it to the following construction problem: "Build a function $f(x)$ that is equal to 1 if the object $a$ satisfies $P(x)$ and equal to 0 if the object $a$ does not satisfy $P(x)$." Then given an object $a$, we compute the value $f(a)$ and know whether $a$ satisfies $P(x)$ or not.

Proposition is proved.

**Remark 2.1.** Not all such reductions are constructive.

**Proposition 2.2.** There is a partial reduction of preservation problems to construction problems.

Proof. Preservation problem reduction:

If we have a preservation problem $Q$, it means that we have to preserve some object $x$. It is possible (at least, in a theoretical setting) to change it to the following construction problem, which includes preservation in a simplified form: "Preserve a description of $x$ and then if necessary, reconstruct $x$ from its description."

Proposition is proved.

These results show that in a theoretical setting, it is possible to consider only construction problems. A paradigmatic constructive problem has the following form:

Given a predicate $P(x)$, build an object $x$ such that $P(x)$ is true.

For algorithms and computer programs, to build means to compute. Although there are two other modes of computer functioning: acceptance and decision [9], they are both reducible to computation. Acceptance is equivalent to computation of a relevant partial indication function. Decision is equivalent to computation of a relevant indication function. Thus, constructive problems are basic in computation theory and computer science.

**3. Recursive Problem Complexity**

In this section, we consider the classical algorithmic complexity for arbitrary problems, restricting ourselves to construction problems for such objects as words in some fixed but arbitrary finite alphabet $A$. Thus, in a general case, we have two types of problems:

(A) Given a predicate P($x$) and a Turing machine $T$, the problem is to compute a word $w$ such that P($w$) is true.

The problem (A) is denoted by $Pr$(P($x$), $T$).

(B) Given a predicate P($x$), the problem is to compute by a Turing machine a word $w$ such that P($w$) is true.

The problem (B) is denoted by $Pr$(P($x$), ***T***) where ***T*** is the class of all Turing machines.

Let us consider some examples.

**Example 3.1.** $P_w$($x$) means "$x$ is a word". In this case, the problem of the second type is to compute some word using a Turing machine.

**Example 3.2.** $P_{new}$($x$) means "$x$ is a non-empty word". In this case, the problem of the second type is to compute some non-empty word using a Turing machine.

**Example 3.3.** P($x = u$) means "$x$ is equal to a word $u$". In this case, the problem of the second type is to compute the word $u$ using a Turing machine.

Any order of symbols in an alphabet $A$, induces lexicographic order in the set $A^*$ of all words in $A$ [10]. This order gives us two more examples of useful predicates that are utilized to define natural algorithmic problems.

**Example 3.4.** P($x \leq z$) means "the word $x$ is less than or equal to the word $z$".

**Example 3.5.** P($x \geq z$) means "the word $x$ is larger than or equal to the word $z$ ".

Finding a given word in a text is one of the most popular search problems. It is formalized in the following example.

**Example 3.6.** P$_{zdiv}$($x$) means "$\exists$ a word $u$ $\exists$ a word $v$ ( $x = uzv$ ) ". The problem is to find a word/text $x$ using a Turing machine such that $x$ contains the word $z$. Usually, it is assumed that such a word $x$ belongs to some specified domain $X$, e.g., to all text on the Internet or to all papers in some journal.

Finding a text with a given word (or words) is one of the most popular search problems on the Internet. It is formalized in the following example.

**Example 3.7.** P$_{divz}$($x$) means "$\exists$ a word $p$ $\exists$ a word $q$ ( $z = pxq$ )". The problem is to find a word/text $x$ using a Turing machine such that $x$ belongs to the word/text $z$. Usually, it is assumed that such a word $x$ satisfies some additional properties, e.g., $x$ is a name of a (given) journal.

Let $l(w)$ denote the length of the word $w$ and $T$ is a Turing machine.

**Definition 3.1.** The *algorithmic/Kolmogorov complexity* or simply, *problem complexity* C{$Pr$(P($x$), $T$)} of the problem $Pr$(P($x$), $T$) relative to the Turing machine $T$ is defined by the following formula:

$$C\{Pr(P(x), T)\} = \min \{ l(p); T(p) = w \text{ and } P(w) \text{ is true}\}.$$

When the machine $T$ does not produce a word $w$ for which P($w$) is true, we put C{$Pr$(P($x$), $T$)} = $\infty$. It is necessary to remark that it also is possible to assume that C{$Pr$(P($x$), $T$)} is not defined when $T$ does not produce a word $w$ for which P($w$) is true.

**Example 3.8.** It is natural to consider Turing machines that, similar to inductive Turing machines, produce their result in a special output tape. Then C$_T$(P$_w$($x$)) = 0 for any Turing machine $T$ that starts working with an empty tape and ends in a final state.

If **P** is a set of predicates on words, then Definition 3.1 determines the function C$^P${$Pr$(P($x$), $T$)} of the variable P with the domain **P**.

If the conventional algorithmic/Kolmogorov complexity C$_T$($x$) of $x$ is interpreted as the length of the shortest program that computes $x$ by means of $T$, the problem complexity C{$Pr$(P($x$), $T$)} is respectively interpreted as the length of the shortest program that solves the problem $Pr$(P($x$)) by means of $T$.

Let P(x) and Q(x) be two predicates on words.

**Proposition 3.1**. If P(x) implies Q(x), then $C\{Pr(Q(x), T)\} \leq C\{Pr(P(x), T)\}$ for any Turing machine $T$.

Indeed, if $C\{Pr(P(x), T)\} = n$ for some element $a$, then there is $p$ such that $T(p) = a$, P($a$) is true, and $l(p) = n$. However, P(x) implies Q(x). So, Q($a$) also is true. Consequently, the shortest input for $T$ that gives an output for which Q(x) is true has the length not larger than $n$, i.e., $C\{Pr(Q(x), T)\} \leq C\{Pr(P(x), T)\}$.

Let us use Proposition 3.1 to obtain properties of problem complexities for some concrete problems.

$C\{Pr(x = z), T)\} = C_T(z)$ for any Turing machine $T$. In such a way, we have the classical theory as a special case of a new one.

**Corollary 3.1**. a) $C\{Pr(x \leq z), T)\} \leq C_T(z)$ for any Turing machine $T$.

b) $C\{Pr(x \leq z), T)\} \leq C\{Pr(x < z), T)\}$ for any Turing machine $T$.

**Proposition 3.2.** a) For any Turing machine $T$, $C\{Pr(x \leq z), T)\} \leq C_T(a)$ for any where $a \leq z$.

b) $C\{Pr(x \leq z), T)\}$ is a non-increasing function, which stabilizes.

c) $C\{Pr(x \leq z), T)\} = \min \{ C_T(y); y \leq z \}$.

Properties of the problem complexity $C\{Pr(P(x), T)\}$ can be different from the properties of the conventional algorithmic/Kolmogorov complexity $C_T(x)$ because they depend not only on $T$ but also on the predicate P(x).

Taking a sequence $\mathbf{P} = \{ P_n(x); n = 1, 2, 3, \ldots \}$ of predicates, we can consider computability and decidability of $C^\mathbf{P}\{Pr(P_n(x), T)\}$ as a function of $n$. Computability and decidability of $C^\mathbf{P}\{Pr(P_n(x), T)\}$ depends on properties both of the predicate P(x) and machine $T$ as the following examples demonstrate. Here we consider computability and decidability only with respect to Turing machines and computability means that we can compute the value $C^\mathbf{P}\{Pr(P_n(x), T)\}$ for all predicates from a set $\mathbf{P}$.

**Example 3.9.** Let $P_n(x) = \{P_w(x) \,\&\, l(x) = n \}$ for all $n = 1, 2, 3, \ldots$ and $T$ is a Turing machine that computes the identity function $e(x) = x$. In this case, $C^\mathbf{P}\{Pr(P_n(x), T)\} = n$, i.e., the value $C\{Pr(P_n(x), T)\}$ is both computable and decidable. At the same time, if the alphabet of Turing machines consists of one symbol and $U$ is a universal

Turing machine, then C{$Pr(P_n(x), U)$} is the classical Kolmogorov/algorithmic complexity, which is not computable (cf., for example, [29]).

**Example 3.10.** It is possible to introduce the lexicographic order on the set of all words in some alphabet. This is a total order. Let P(x) be an undecidable predicate such that taking any number $n$ and all words with the length $n$, P(x) can be true only for the largest of these words, $P_n(x) = P(x) \& l(x) = n$ and $T$ is a Turing machine that computes the identity function $e(x) = x$. In this case, if we can decide whether C{$Pr(P_n(x), T)$} = $n$, then we can decide whether P(x) is true for any word with the length $n$. As P(x) is an undecidable predicate, C{$Pr(P(x), T)$} also is undecidable.

**Example 3.11.** Let us consider the set of predicates $\mathbf{P}_A$ = {P($x = u$); $u$ is a word in an alphabet $A$, i.e., $u \in A^*$} (cf. Example 3.3). In this case, the function C{$Pr(P(x = u), T)$} is the classical Kolmogorov/algorithmic complexity $C_T(u)$ relative to the Turing machine $T$. This function is not computable when $T$ is a universal Turing machine (cf., for example, [40]).

**Definition 3.2.** A function $f(n)$ is called *additively smaller* than a function $g(n)$ if there is such a number $k$ that $f(n) \leq g(n) + k$ for all $n \in N$.

The relation $f(n) \leq g(n) + k$ for all $n \in N$ is denoted by $f(n) \preccurlyeq g(n)$.

Let **H** be a class of functions.

**Definition 3.3.** A function $f(n)$ is called *additively optimal* for the class **H** if it is additively smaller than a function from **H**, i.e., there is such a number $k$ that $f(n) \leq g(n) + k$ for any $g \in \mathbf{H}$ and all $n \in N$.

**Remark 3.1**. Additive optimality is a special case of a general functional optimality introduced in [5]. In particular, there are other kinds of optimality, e.g., multiplicative optimality where $f(n) \leq k \cdot g(n)$.

**Definition 3.4.** Functions $f(n)$ and $g(n)$ are called *additively equivalent* if $f(n) \preccurlyeq g(n)$ and $g(n) \preccurlyeq f(n)$.

This relation is denoted by $f(n) \asymp g(n)$.

**Proposition 3.3** [5]. Any two functions additively optimal for a class **H** are additively equivalent if they both belong to the class **H**.

Proposition 3.3 shows that additively optimal functions in general and dual measures, in particular, are in some sense invariant.

In the theory of Kolmogorov complexity, it is proved that there are optimal elements (cf., for example, [40]).

Although there are much more predicates than those that are in the class $\mathbf{P}_A = \{P(x = u); u \in A^*\}$, the class of problem complexities has optimal elements.

**Definition 3.5.** A problem complexity $C\{Pr(P(x), M)\}$ is called *additively optimal* for a set $\mathbf{P}$ of predicates on words if it is additively optimal for all functions $C\{Pr(P(x), T)\}$ with $P \in \mathbf{P}$ and $T \in \mathbf{T}$, i.e., for any Turing machine $T$ there is a number $k$ such that for any predicate P from the class $\mathbf{P}$, we have $C\{Pr(P(x), M)\} \leq C\{Pr(P(x), T)\} + k$.

Let $\mathbf{P}$ be an arbitrary set of predicates on words.

**Theorem 3.1.** For any set $\mathbf{P}$ of predicates on words, there is an additively optimal problem complexity $C\{Pr(P(x), M)\}$, i.e., for any Turing machine $T$ there is a number $k$ such that for any predicate $P(x)$ from the class $\mathbf{P}$, we have $C\{Pr(P(x), M)\} \leq C\{Pr(P(x), T)\} + k$.

Proof. Let us consider a Turing machine $T$, a predicate $P(x)$ from $\mathbf{P}$, and a universal Turing machine $U$, which given the word $<w, \mathbf{c}(T)>$ as the input produces the same result as $T$ given the word $w$ as the input, i.e., $U(<w, \mathbf{c}(T)>) = T(w)$ (cf., for example, [46]). It is possible to build a pairing function $<>$ with the following property: $l(<w, u>) = l(w) + k_u$ where $k_u$ depends only on the word $u$, i.e., for a fixed word $u$, the number $k_u$ is the same for all words $w$.

For the value $C\{Pr(P(x), T)\}$, we have two possibilities: either for some word $z$, the predicate $P(T(z))$ is true, or there is no such word $z$ for which the predicate $P(T(z))$ is true. In the first case, we can take word $z$ such that $C\{Pr(P(x), T)\} = l(z)$. Then $T(z) = U(<z, \mathbf{c}(T)>)$ and $P(U(<z, \mathbf{c}(T)>))$ is true. Thus, $C\{Pr(P(x), U)\} \leq l(<z, \mathbf{c}(T)>) = l(z) + k_T = C\{Pr(P(x), T)\} + k_T$, i.e., the inequality that we need to prove is true.

In the second case, $C\{Pr(P(x), T)\} = \infty$ and $C\{Pr(P(x), U)\} \leq C\{Pr(P(x), T)\} + k_T$ because $\infty$ is larger than any number.

Theorem is proved because the constant $k_T$ does not depend on the predicate $P(x)$.

Taking $\mathbf{P} = \mathbf{P}_A = \{P(x = u); u \in A^*\}$, we have the classical result.

**Corollary 3.2** (Kolmogorov, Chaitin). In the set of the classical Kolmogorov/algorithmic complexities $C_T(x)$, there is an additively optimal complexity $C(x)$, for which $C(x) \leq C_T(x) + k$ for any Turing machine $T$.

Thus, Theorem 3.1 means that for any set of predicates, the set of problem complexities has complexities $C\{Pr(P(x), U)\}$ that are invariant up to equivalence with respect to changing universal Turing machines. This justifies the following definition.

**Definition 3.6.** The *recursive algorithmic/Kolmogorov complexity* or simply, *recursive problem complexity* $C\{Pr(P(x))\}$ of the problem $Pr(P(x), T)$ is equal to $C\{Pr(P(x), U)\}$ where $U$ is a universal Turing machine.

By Definition 3.1, we have $C\{Pr(P(x))\} = \min \{ l(p); U(p) = x$ and $P(x)$ is true$\}$ where $U$ is a universal Turing machine. In other words, by Theorem 3.1, recursive problem complexity $C\{Pr(P(x))\}$ is additively optimal in the class of all problem complexities with respect to Turing machines.

Proposition 3.1 implies the following result.

**Corollary 3.3.** If $P(x)$ implies $Q(x)$, then $C\{Pr(Q(x))\} \leq C\{Pr(P(x))\}$.

Properties of the problem complexity $C\{Pr(P(x))\}$ can be different from the properties of the conventional algorithmic/Kolmogorov complexity $C(x)$ because problem complexity depends on the class of predicates **P**. For instance, it is proved that $C(x)$ is not computable (cf., [40]). At the same time, we can find a sequence **P** = { $P_n(x)$; $n = 1, 2, 3, \ldots$ } of predicates for which the problem complexity $C\{Pr(P_n(x), U)\}$ is not only computable but also decidable.

**Example 3.12.** Let us take a sequence **P** = { $P_n(x)$; $n = 1, 2, 3, \ldots$ } of predicates $P_n(x)$ where $P_n(x)$ means "the word $x$ is computed by a chosen universal Turing machine $U$ in less than $n$ steps." Then the function $C\{Pr(P_n(x))\} = C\{Pr(P_n(x), U)\}$ is computable and even decidable with respect to $n$ because it is possible to check if $U$ gives a result making $n$ steps or less and there are such inputs for which $U$ gives a result making $n$ steps or less as $U$ simulates a Turing machine that computes the identity function.

Moreover, it is possible that the function $C\{Pr(P_i(x))\}$ is constant.

**Example 3.13.** To show this, let us take a finite alphabet $A$ and a sequence **P** = $\{P_z(x) = P(x < z)$; $z$ is a non-empty word in the alphabet $A$ $\}$ of predicates $P_z(x)$ where P($x < z$) means "the word $x$ is less than the word $z$." Then the function C{$Pr(P_z(x))$} = C{$Pr(P_i(x), U)$} is equal to 1 for the following universal Turing machine $U$. The machine $U$ simulates Turing machines so that codes of Turing machines are their numbers in some (Gödel) enumeration. In addition, the first Turing machine $T_1$ in this enumeration outputs the least non-empty word given the empty input, i.e., when nothing is written in the tape of this machine at the beginning. Changing the universal Turing machine $U$, we can make the function C{$Pr(P_z(x))$} = C{$Pr(P_i(x), U)$} equal to any natural number.

Optimal algorithmic complexities of words allow one to solve many important problems in the theory of algorithms and to build constructive probability theory and algorithmic information theory. However, for arbitrary predicates, properties of optimal problem complexity can essentially depend on the choice of the universal Turing machine $U$ as the following results show.

**Proposition 3.4.** For any natural number $n > 0$, there is a universal Turing machine $U$ such that C{$Pr(P_z(x))$} = C{$Pr(P_i(x), U)$} = $n$.

Proof. This result is proved for $n = 1$ in Example 3.13. So, we may assume that $n > 1$. It is possible to build a pairing function <>: $(A^*) \times (\mathbf{c}T) \to A^*$ such that all words from $A^*$ with length less than $n$ are equal to words <$\varepsilon$, $\mathbf{c}(T_k)$> where $T_k$ are Turing machines that produce no outputs given the empty input. In addition, the least word with $n$ symbols is equal to a word <$\varepsilon$, $\mathbf{c}(T)$> where $T$ is a Turing machine that produces the least non-empty word $x$ as its output given the empty input. The universal Turing machine $U$ that works with this pairing function determines the problem complexity C{$Pr(P_z(x))$} = C{$Pr(P_i(x), U)$} = $n$ as $x$ is the least non-empty word in $A^*$.

Proposition is proved.

**Remark 3.2.** The equality $C_U(P(x)) = a$ is, in general, undecidable when P($x$)) is an undecidable predicate.

At the same time, choosing another universal Turing machine, we can make problem complexity undecidable even for very simple problems.

Let us consider all Turing machines that work with words in an alphabet $A$.

**Proposition 3.5**. The equality C{$Pr(P_w(x), U)$} = 1 is undecidable with respect to some universal Turing machine $U$.

Proof. Let us assume that all words in the alphabet $A$ are ordered and written in a sequence $x_1, x_2, \ldots, x_n, \ldots$ and $l(x_1) = 1$. Then for any word $x$ in the alphabet $A$ and any Turing machine $T$, it is possible to build a Turing machine $V_{T,x}$ such that $V_{T,x}(x_1) = T(x)$. In addition, for any Turing machine $V$, it is possible to build a universal Turing machine $U(V)$ such that $U(V)(x_1) = V(x_1)$ and $U$ does not halt for all other words of length one. Consequently, $C\{Pr(P_w(x), U)\} = 1$ if and only if $V(x_1)$ is defined. When $V = V_{T,x}$, the value $V_{T,x}(x_1)$ is defined if and only if the machine $T$ gives the result for input $x$. However, this gives us the halting problem for Turing machines and this problem is undecidable. Previous considerations reduce the halting problem to the problem of $C\{Pr(P_w(x), U)\} = 1$ being equal to one. Consequently, the equality $C\{Pr(P_w(x), U)\} = 1$ also is undecidable.

Proposition is proved.

It shows that properties of the problem complexity can depend on the chosen universal Turing machine $U$. This feature of problem complexity contrasts to the properties of the conventional algorithmic/Kolmogorov complexity, which is in some sense invariant (cf., for example, [40]).

Any universal Turing machine $U$ computes all words. It implies the following result.

**Proposition 3.6.** The function $C\{Pr(P(x))\}$ is defined if and only if $P(x)$ is consistent, i.e., $P(x)$ is not identically false.

Theorem 3.1 implies the following result.

**Proposition 3.7.** Any two additively optimal problem complexities for any set **P** of predicates are additively equivalent.

**Corollary 3.4.** Any two recursive problem complexities are additively equivalent for any set of predicates **P**.

**Corollary 3.5** (Kolmogorov, Chaitin). Any two optimal problem complexities are additively equivalent for the set of predicates $\mathbf{P}_A$.

Let $\mathbf{P} = \{P_n(x); n = 1, 2, 3, \ldots\}$ be a sequence of predicates.

**Theorem 3.2.** The problem complexity $C\{Pr(P_n(x)), M)\}$ tends to infinity when $n$ tends to infinity if and only if for any finite subset $D$ from $A^*$ there is a number $n$ such that there is no $d$ in $D$ for which any predicate $P_m(x)$ is true when $m > n$.

Proof. Sufficiency. Let us take the set *D* that consists of all words in the alphabet *A* that have length less than *k*. Then by the condition of the theorem, there is a number *n* such that the predicate $P_n(x)$ can be true only when the length of the word *x* is equal to or larger than *k*. Consequently, $C\{Pr(P_n(x)), M\} \geq k$. Thus, $C\{Pr(P_n(x)), M\}$ tends to infinity when *n* tends to infinity because we can take an arbitrary number *k* when we build the set *D*.

Necessity. Let *D* be a finite subset of $A^*$ and $k = \max\{l(x); x \in D\}$. If $C\{Pr(P_n(x)), M\}$ tends to infinity when *n* tends to infinity, then there a number *n* such that $C\{Pr(P_m(x)), M\} > k$ for all $m > n$. At the same time, if $P_m(x) \leq k$ for some $m > n$, then $C\{Pr(P_m(x)), M\} \leq k$. Consequently, there is no *d* in *D* for which any predicate $P_m(x)$ is true when $m > n$.

Theorem is proved.

**Condition (F).** For any subset *D* from $A^*$ the set $\{n; \exists d \in D\ (P_n(d)\ \text{is true})\}$ is finite.

**Corollary 3.5.** If the condition (F) is true for predicates **P** = { $P_n(x)$; $n$ = 1, 2, 3, … }, then the problem complexity $C\{Pr(P_n(x)), M\}$ tends to infinity when *n* tends to infinity.

Indeed, the condition (F) implies that the minimal length of words for which $P_n(d)$ is true grows with *n*. Any Turing machine transforms any finite number of words into a finite number of words. Thus, the formula $C\{Pr(P_n(x))\} = \min \{ l(p); U(p) = x$ and $P_n(x)$ is true$\}$ implies that minimal length of words that are transformed by *U* into words for which $P_n(d)$ is true grows without limits when *n* tends to infinity.

**Corollary 3.6.** If for any number *n*, there is a number *m* such that the truth of $P_n(x)$ implies that $l(x) > m$ and *m* tends to infinity when *n* tends to infinity, then the values of the problem complexity $C\{Pr(P_n(x)), M\}$ tends to infinity when *n* tends to infinity.

Corollaries 3.5 and 3.6 imply the following classical result.

**Corollary 3.7** (Kolmogorov, Chaitin)**.** Algorithmic complexity $C(x)$ tends to infinity when $l(x)$ tends to infinity.

This result shows that there are many cases when the problem complexity has similar properties to properties of algorithmic complexity. Another property of algorithmic complexity is that this function is not monotone [40]. Thus, it is natural to ask the following question:

Is it possible to find a problem complexity that is monotone?

The following example gives a positive solution to this question.

**Example 3.14.** Let us take a sequence $\mathbf{P} = \{ P_n(x); n = 1, 2, 3, \ldots \}$ of predicates $P_n(x)$ where $P_n(x)$ means "the algorithmic complexity $C(x)$ of the word $x$ is equal to $n$." It means that taking a universal Turing machine $U$, there is a word $z$ for which $U(z) = x$, $l(z) = n$, and for all shorter words $U$ does not give $x$ as its output. Then the function $C\{Pr(P_n(x))\} = C\{Pr(P_n(x), U)\} = n$ for all $n$. In addition, this function is decidable with respect to $n$.

It is interesting that although all predicates $P_n(x)$ are not decidable, and even not semi-decidable, their problem complexity is decidable.

Let us consider two classes of predicates $\mathbf{P} = \{ P_i(x); i \in I \}$ and $\mathbf{Q} = \{ Q_i(x); i \in I \}$ such that there is a Turing machine $T_{\mathbf{PQ}}$ that given a word $w$ that satisfies a predicate $P_i(x)$, computes a word $v$ that satisfies the predicate $Q_i(x)$.

**Theorem 3.3.** There is a number $k$ such that $C\{Pr(Q_i(x))\} \leq C\{Pr(P_i(x))\} + k$ for all $i \in I$.

**Proposition 3.8.** a) For any universal Turing machine $U$, $C\{Pr((x \geq z), U)\} \to \infty$ when $z \to \infty$.

b) The function $C\{Pr((x \geq z), U)\}$ is a not computable by a Turing machine.

c) $C\{Pr((x \geq z), U)\} = \min \{ C_T(y); y \geq z \}$.

Proposition 3.8 shows that the function $C\{Pr(x \geq z)\}$ is equal to the function $mC(z)$, which is often used in the theory of Kolmogorov complexity (cf., [8, 40]).

It is useful to consider problems with respect to sets of predicates because some important axiom systems use axiom schemas, which define sets of predicates. Examples of such schemas are the Replacement Axiom and Axiom of Subsets in Zermelo-Fraenkel set theory [28]. This introduces a new kind of constructive problems:

(C)     Given a set of predicates $\mathbf{P}$ and a Turing machine $T$, compute a word $w$ such that $P(w)$ is true for any predicate $P(x)$ from $\mathbf{P}$.

The problem (A) is denoted by $Pr(\mathbf{P}, T)$.

**Definition 3.7.** The *algorithmic/Kolmogorov complexity* or simply, *problem complexity* C{$Pr(\mathbf{P}, T)$} of the problem $Pr(\mathbf{P}, T)$ relative to the Turing machine $T$ is given by the following formula:

C{$Pr(\mathbf{P}, T)$} = min { $l(p)$; $T(p) = w$ and P($w$) is true for all predicates P($x$) from **P** }.

When $T$ does not produce a word $w$ for which P($w$) is true for all predicates P($x$) from **P**, then we put C{$Pr(\mathbf{P}, T)$} = ∞.

Let **P** and **Q** be two sets of predicates on words.

**Proposition 3.9**. If for any predicate Q($x$) from **Q**, there is a predicate P($x$) from **P** such that P($x$) $\Rightarrow$ Q($x$), i.e., P($x$) implies Q($x$), then C{$Pr(\mathbf{Q}, T)$} $\leq$ C{$Pr(\mathbf{P}, T)$} for any Turing machine $T$.

### 4. Problem Complexity with respect to classes of algorithms

The initial Kolmogorov or algorithmic complexity was defined by its creators for the class of all Turing machines because this class was at that time believed to be an absolute class that comprises up to equivalence all other classes. This provided for a belief that in such a way we got a universal optimal complexity, which up to some additive constant gave the least complexity for semiotic objects. This was an attempt to build a universal dual complexity measure, which does not depend on a specific class of algorithms. However, this goal has not been achieved. One reason was that it turned out that the original definition was not sufficient for solving some mathematical and practical problems. For example, such universal measure was not appropriate for formalization of the concept of randomness and for the development of algorithmic probability theory and information theory. The second reason for impossibility to achieve this goal (and for necessity of constructing relative dual measures) was the discovery of super-recursive algorithms. Before it happened, all believed that Turing machines or other class of recursive algorithms give an absolute, universal model for algorithms and computation. Emergence of super-recursive algorithms changed the situation. In the universe of super-recursive algorithms, there are no absolutely universal classes or models. The third reason for impossibility to build a universal dual complexity measure was that actually computer scientists have already used several distinct dual measures. As a result, the universal approach was discarded and it has become necessary to define complexity relative to a given class of algorithms. At first, it was done for some specific classes like monotone Turing machines or prefix partial recursive functions. Then dual measures have been introduced and studied [5]. Later an axiomatic approach to dual complexity measures has been elaborated [10].

Let $\mathbf{A} = \{ A_i ; i \in I\}$ be a class of algorithms that work with words in an alphabet $X$. To characterize complexity of objects with respect to a class of algorithms, we take optimal measures.

**Definition 4.1**[5]**.** The *Kolmogorov/algorithmic complexity* $C_\mathbf{A}(x)$ of an object/word $x$ with respect to the class $\mathbf{A}$ is defined as

$$C_\mathbf{A}(x) = \min \{ l(p); \ U(p) = x\}$$

where $l(p)$ is the length of the word $p$ and $U$ is a universal algorithm in the class $\mathbf{A}$.

When the algorithm $U$ does not produce the word $x$, and it means that no algorithm in $\mathbf{A}$ can do this, we define $C_\mathbf{A}(x) = \infty$.

We remind that an algorithm $U$ is called *universal* for the class $\mathbf{A}$ if for any algorithm $A$ from $\mathbf{A}$ and any word $x$, the word $p = <\mathbf{c}(A), x>$ is given as the input to $A$, the result of $U$ is equal to the result of $A$ applied to $x$ [5, 10]. Examples of universal algorithms are a universal Turing machine and a universal inductive Turing machine [7].

The dual complexity measure that corresponds to a universal algorithm gives an invariant characteristic of the whole class $\mathbf{A}$.

Kolmogorov/algorithmic complexity $C_\mathbf{A}(x)$ is a particular case of static dual complexity measures considered in [10]. At the same time, many other complexity measures studied by different authors are special cases of complexity $C_\mathbf{A}(x)$:

When Kolmogorov complexity is defined for the class of Turing machines that compute symbols of a word $x$, we obtain uniform complexity $KR(x)$ studied by Loveland [41].

When Kolmogorov complexity is defined for the class of prefix functions, we obtain prefix complexity K(*x*) studied by Gacs [29] and Chaitin [16].

When Kolmogorov complexity is defined for the class of monotone Turing machines, we obtain monotone complexity Km(*x*) studied by Levin [38].

When Kolmogorov complexity is defined for the class of Turing machines that have some extra initial information, we obtain conditional Kolmogorov complexity CD(*x*) studied by Sipser [51].

Let *t*(*n*) and *s*(*n*) be some functions of natural number variables.

When Kolmogorov complexity is defined for the class of recursive automata that perform computations with time bounded by some function of a natural variable *t*(*n*), we obtain time-bounded Kolmogorov complexity $C^t(x)$ studied by Kolmogorov [35] and Barzdin [2].

When Kolmogorov complexity is defined for the class of recursive automata that perform computations with space (i.e., the number of used tape cells) bounded by some functions of a natural variable *s*(*n*), we obtain space-bounded Kolmogorov complexity $C^s(x)$ studied by Hartmanis and Hopcroft [31].

When Kolmogorov complexity is defined for the class of multitape Turing machines that perform computations with time bounded by some function *t*(*n*) and space bounded by some function *s*(*n*), we obtain resource-bounded Kolmogorov complexity $C^{t,s}(x)$ studied by Daley [18].

Quantum Kolmogorov complexity [54] also is a special case of the dual complexity measure $C_A(x)$:

All of these kinds of complexity are dual complexity measures. The generalized Kolmogorov complexity introduced and studied in [5, 8] gives a general setting for all of them.

However, here we are interested in problem complexity defined for a wide variety of problems.

Let P(*x*) be a predicate on words in the alphabet *A* and *Pr*(P(*x*)) be a problem of finding/constructing/computing such a word that satisfies P(*x*) by algorithms from **A**.

**Definition 4.2.** The *algorithmic/Kolmogorov complexity* or simply, *problem complexity* $C_A\{Pr(P(x))\}$ of the problem $Pr(P(x))$ with respect to the class **A** is given by the following formula:

$$C_A\{Pr(P(x))\} = \min \{ l(p); U(p) = w \text{ and } P(w) \text{ is true}\}$$

where $U$ is a universal algorithm in the class **A**.

When the algorithm $U$ does not produce a word $w$ for which $P(w)$ is true, then we put $C_A\{Pr(P(x))\} = \infty$. It is necessary to remark that it also is possible to assume that $C_A\{Pr(P(x))\}$ is not defined when $U$ does not produce a word $w$ for which $P(w)$ is true.

Let the class **A** contains an identity algorithm $E$ that computes the function $e(x) = x$.

**Proposition 4.1.** A problem $Pr(P(x))$ has a solution, i.e., there is a word that satisfies $P(x)$, if and only if the value $C_A\{Pr(P(x))\}$ is a natural number.

The length of a word/text $x$ is, according to the general theory of information [6], a kind of a measure of information in this word/text $x$. Thus, the problem complexity $C_A\{Pr(P(x))\}$ estimates minimal information necessary to solve the problem $Pr(P(x))$ by algorithms from the class **A**. In particular, for the predicate $x = z$, the function $C_A\{Pr(x = z)\}$ estimates minimal information necessary to build (compute) $z$ by algorithms from **A**. This is the classical Kolmogorov complexity. As we see, it estimates not the information in the word $z$, as some assert, but information necessary to compute (get) $z$, i.e., information about $z$. It is definitely different problems to get information how to get $z$ and information that is contained in $z$. For example, if you have a coded text $T$, information how to get or build $T$ is zero because you already have $T$. However, complexity of getting information from $T$ may be very large.

If **P** is a set of predicates on words, then $C_A^P\{Pr(P(x), T)\}$ is a function of the variable P with the domain **P**.

Let us consider a class of predicates **P** and two classes of algorithms **A** and **B** that work with words in alphabets $X$ and $Z$, correspondingly.

**Proposition 4.2.** If $X \subseteq Z$ and for any algorithm $A$ from **A** there is an algorithm $B$ from **B** such that for any $x \in X$ either $B(x) = A(x)$ or $A$ is undefined for $x$, then $C_B^P\{Pr(P(x))\} \leqslant C_A^P\{Pr(P(x))\}$ for any predicate P from **P**.

This result brings forth hierarchies of problem complexities.

Let $\mathbf{IM}_n$ be a class of inductive Turing machines of order $n$ [7]. Then we have the following hierarchy:

$$C_T{}^\mathbf{P}\{Pr(P(x))\} \preccurlyeq C_{\mathbf{IM}_1}{}^\mathbf{P}\{Pr(P(x))\} \preccurlyeq C_{\mathbf{IM}_2}{}^\mathbf{P}\{Pr(P(x))\} \preccurlyeq \ldots \preccurlyeq C_{\mathbf{IM}_n}{}^\mathbf{P}\{Pr(P(x))\} \preccurlyeq \ldots$$

Results from [8] show that this hierarchy is proper, i.e., all inequalities are strict, when $P(x)$ is $z = x$.

**Remark 4.1**. It is possible to chose universal algorithms so that we obtain inequality of functions $C_\mathbf{B}{}^\mathbf{P}\{Pr(P(x))\} \leq C_\mathbf{A}{}^\mathbf{P}\{Pr(P(x))\}$, but as the function itself is defined up to equivalence, in a general case, it is possible to assert only relation $\preccurlyeq$.

**Corollary 4.2.** If $X \subseteq Z$ and $\mathbf{A} \subseteq \mathbf{B}$, then $C_\mathbf{B}{}^\mathbf{P}\{Pr(P(x))\} \preccurlyeq C_\mathbf{A}{}^\mathbf{P}\{Pr(P(x))\}$ for any predicate P from $\mathbf{P}$.

**Remark 4.2**. Taking general Turing machines and other classes of algorithms studied in [48], it possible to build problem complexity for these classes. Then Proposition 4.1 determines a hierarchy of problem complexities similar to the hierarchy of Kolmogorov complexities constructed in [48].

**Remark 4.3**. Even if there is a proper inclusion of classes $\mathbf{A} \subset \mathbf{B}$, it does not mean the strict inequality $C_\mathbf{B}\{Pr(P(x))\} \prec C_\mathbf{A}\{Pr(P(x))\}$. We can take as an example the pair of classes $\mathbf{T}$ of all Turing machines and $\mathbf{U}$ of all universal Turing machines.

Let $\mathbf{P}$ be an arbitrary set of predicates on words and the class $\mathbf{A}$ has a universal algorithm.

**Theorem 4.1.** For any set $\mathbf{P}$ of predicates on words, there is an additively optimal problem complexity $C\{Pr(P(x), M)\}$, i.e., for any algorithm H from the class $\mathbf{A}$ there is a number $k$ such that for any predicate $P(x)$ from the class $\mathbf{P}$, we have $C\{Pr(P(x), M)\} \leq C\{Pr(P(x), T)\} + k$ or $C^\mathbf{P}\{Pr(P(x), M)\} \leq C^\mathbf{P}\{Pr(P(x), T)\} + k$.

Proof. Let us consider an algorithm H from the class $\mathbf{A}$, a predicate $P(x)$ from $\mathbf{P}$, a codification $\mathbf{c}: \mathbf{A} \to A^*$ of algorithms from $\mathbf{A}$, and a universal in $\mathbf{A}$ algorithm U, which given the word $<w, \mathbf{c}(H)>$ as the input produces the same result as H given the word $w$ as its input, i.e., $U(<w, \mathbf{c}(H)>) = H(w)$ (cf., for example, [46]). It is possible to build a pairing function $<>$ with the following property: $l(<w, u>) = l(w) + k_u$ where $k_u$ depends only on the word $u$, i.e., for a fixed word $u$, the number $k_u$ is the same for all words $w$.

For the value $C\{Pr(P(x), H)\}$, we have two possibilities: either for some word $z$, the predicate $P(H(z))$ is true, or there is no such word $z$ for which the predicate $P(H(z))$ is true. In the first case, we can take word $z$ such that $C\{Pr(P(x), H)\} = l(z)$. Then $H(z) = U(<z, \mathbf{c}(H)>)$ and $P(U(<z, \mathbf{c}(H)>))$ is true. Thus, $C\{Pr(P(x), U)\} \leq l(<z, \mathbf{c}(H)>) = l(z) + k_H = C\{Pr(P(x), H)\} + k_H$, i.e., the inequality that we need to prove is true.

In the second case, $C\{Pr(P(x), H)\} = \infty$ and $C\{Pr(P(x), U)\} \leq C\{Pr(P(x), H)\} + k_H$ because $\infty$ is larger than any number.

Theorem is proved because the constant $k_H$ does not depend on the predicate $P(x)$.

The result of Theorem 4.1 spares a researcher and a student to prove optimality for problem complexity with respect to different classes of algorithms. The result of Proposition 3.2 shows that up to additive optimality, recursive problem complexity is invariant, i.e., it is independent of the choice of universal Turing machine.

A theory of problem complexity with respect to classes of algorithms can be developed parallel to presented in Section 3 theory of problem complexity with respect to Turing machines. For instance, it was developed for a problem related to the predicate $x = z$ and for the class of all inductive Turing machines of the first order in [8]. Axiomatic theory of algorithms [9] gives the best context for the development of a theory for problem complexity with respect to classes of algorithms. We are not doing it here as this work has different goals.

## 5. Functional Problem Complexity

The classical algorithmic complexity $C(x)$ is defined for finite objects such as words. However, it is important and interesting to study algorithmic complexity for infinite objects such as functions, languages or sequences. Algorithms can build/compute such objects. In the classical theory and inductive computations, they do this potentially. For instance, it is assumed that a given Turing machine computes some function or decides some language. Some algorithms are even named as functions, e.g., partial recursive functions. In the theory of hyper-computation, it is assumed that it is possible to completely build/compute infinite objects [10]. Thus, it is important, even for practical purposes, to know complexity of such computations. As languages, sequences and many other infinite objects can be represented by functions, we introduce here algorithmic/Kolmogorov complexity of functions with respect to the class of all Turing machines and then study problem complexity in this class. In what follows a function can be partial.

Let $l(w)$ denote the length of the word $w$, $f: A^* \to A^*$ is a (partial) function, and $T$ is a Turing machine input of which consists of two words in the alphabet $A$, e.g., $A$ has two input tapes.

**Definition 5.1.** The *algorithmic/Kolmogorov complexity* $C\{f(x), T\}$ of the function $f(x)$ relative to the Turing machine $T$ is given by the following formula:

$$C\{f(x), T\} = \min\{l(p); T(p, x) = f(x)\}.$$

When there is no word $p$ such that $T(p, x) = f(x)$, then we put $C\{f(x), T\} = \infty$. It is necessary to remark that it also is possible to assume that $C\{f(x), T\}$ is not defined when there is no word $p$ such that $T(p, x) = f(x)$.

Such kind of algorithmic/Kolmogorov complexity also is considered in [29]. There are other approaches to algorithmic complexity of infinite objects, but we do not consider them here.

**Proposition 5.1.** If $C\{f(x), U\} = \infty$ for a universal Turing machine $U$, then the function $f(x)$ is recursively noncomputable, that is, noncomputable by Turing machines.

Note that while for any word there is a Turing machine that computes this word, there are noncomputable functions. In other words, the value of $C(x) = C\{x, U\}$ is always finite for a universal Turing machine $U$, while $C\{f(x), U\}$ can be infinite.

Algorithmic complexity of functions allows us to define and study problems related to functions, such as: "Build a function with given properties" or "Find whether two functions coincide."

Functional problem complexity, we introduce with respect to an arbitrary class of algorithms **A**. Let us consider a predicate $P(f)$ on functions and the construction problem $Pr(P(f))$ that demands to build/compute a function $f(x)$ such that $P(f(x))$ is true.

**Definition 5.2.** The *algorithmic/Kolmogorov complexity for functions* or simply, *functional problem complexity* $C_\mathbf{A}\{Pr(P(f))\}$ of the problem $Pr(P(f))$ with respect to the class **A** is given by the following formula:

$$C_\mathbf{A}\{f(x)\} = \min\{l(p); U(p, x) = f(x) \text{ and } P(f(x)) \text{ is true}\}$$

where $U$ is a universal algorithm in the class **A**.

When the algorithm $U$ cannot compute a function $f(x)$ for which $P(f)$ is true, then we put $C_A\{f(x)\} = \infty$.

In what follows, we consider construction problems for such objects as functions on words in some fixed but arbitrary finite alphabet $A$. Means of constructions are Turing machines with two input tapes, or in general algorithms with two inputs from some given class. One input is treated as the argument of the computed function. For another input, there are different interpretations: a) a program for computation; b) name/index of the computed function; c) description of the computed function. Thus, in a general case, we have the following problems:

(A)    Given a predicate $P(F)$ on functions where $F$ is a functional variable and a Turing machine $T$, compute a function $f(x)$ such that $P(f(x))$ is true, i.e., it is necessary to find a word $p$ such that $T(p, x) = f(x)$ and $P(f(x))$ is true.

Such problem (A) is denoted by $Pr(P(F)), T)$.

(B)    Given a predicate $P(F)$, compute by some Turing machine a function $f(x)$ such that $P(f(x))$ is true.

Such problem (B) is denoted by $Pr(P(F)), \boldsymbol{T})$ where $\boldsymbol{T}$ is the class of all Turing machines.

The problem (B) is reduced to the problem (A) because a universal Turing machine can compute any function computable by some Turing machine. As it is known [46], these functions form the class of partial recursive functions.

**Definition 5.3.** The *algorithmic/Kolmogorov complexity* or simply, *problem complexity* $C\{Pr(P(F)), T)\}$ of the problem $Pr(P(x), T)$ relative to the Turing machine $T$ is given by the following formula:

$$C\{Pr(P(F)), T)\} = \min\{l(p); T(p, x) = f(x) \text{ and } P(f) \text{ is true}\}.$$

When $T$ does not produce a word $w$ for which $P(w)$ is true, then we put $C\{Pr(P(F), T)\} = \infty$. It is necessary to remark that it also is possible to assume that $C\{Pr(P(F)), T)\}$ is not defined when $T$ does not produce a word $w$ for which $P(w)$ is true.

If **P** is a set of predicates on words, then Definition 5.3 determines the function $C^{\mathbf{P}}\{Pr(P(F), T)\}$ of the variable P with the domain **P**.

Let **P** be an arbitrary set of predicates on words.

**Theorem 5.1.** For any set **P** of predicates on words, there is an optimal problem complexity $C\{Pr(P(F), M)\}$ such that for any Turing machine $T$ there is a number $k$ such that for any predicate $P(F)$ from the class **P**, we have $C^{\mathbf{P}}\{Pr(P(F), M)\} \leq C^{\mathbf{P}}\{Pr(P(F), T)\} + k$.

Proof. Let us consider a Turing machine $T$, a predicate $P(F)$ from **P**, and a universal Turing machine $U$, which given the pair $(<p, \mathbf{c}(T)>, x)$ as the input produces the same result as $T$ given the pair of words $(p, x)$ as the input, i.e., $U(<p, \mathbf{c}(T)>, x) = T(p, x)$. It is possible to build a pairing function $<>$ with the following property: $l(<(p, u>) = l(p) + k_u$ where $k_u$ depends only on the word $u$, i.e., for a fixed word $u$, the number $k_u$ is the same for all words $w$.

For the value $C\{Pr(P(F), T)\}$, we have two possibilities: either for some word $z$, the predicate $P(T(z, x))$ is true, or there is no such word $z$ for which the predicate $P(T(z, x))$ is true. In the first case, we can take word $z$ such that $C\{Pr(P(F), T)\} = l(z)$. Then $T(z, x) = U(<z, \mathbf{c}(T)>, x)$ and $P(U(<(z, x), \mathbf{c}(T)>))$ is true. Thus, $C\{Pr(P(F), U)\} \leq l(<z, \mathbf{c}(T)>) = l(z) + k_T = C\{Pr(P(x), T)\} + k_T$, i.e., the inequality that we need to prove is true.

In the second case, $C\{Pr(P(F), T)\} = \infty$ and $C\{Pr(P(F), U)\} \leq C\{Pr(P(F), T)\} + k_T$ because $\infty$ is larger than any number.

Theorem is proved because the constant $k_T$ does not depend on the predicate $P(F)$.

Taking $\mathbf{P} = \mathbf{P}_z$, we have the classical result.

**Corollary 5.1.** In the set of the classical Kolmogorov/algorithmic complexities $C_T(f)$, there is an optimal complexity $C(f)$, for which $C(f) \leq C_T(f) + k$ for any Turing machine $T$.

The optimal (recursive) algorithmic/Kolmogorov problem complexity $C\{Pr(P(F), U)\}$ in the class of problems $Pr(P(F))$ with $P(F)$ in **P** is denoted by $C\{Pr(P(F))\}$.

By Theorem 3.1, we have $C\{Pr(P(x))\} = \min \{ l(p); U(p) = x$ and $P(x)$ is true$\}$ where $U$ is some universal Turing machine.

When the algorithm $U$ cannot compute a function $f(x)$ for which $P(f)$ is true, then we put $C_{\mathbf{A}}\{f(x)\} = \infty$.

Let us assume that the class **A** is closed under sequential composition and consider two classes of predicates **P** = { $P_i(f)$; $i \in I$ } and **Q** = { $Q_i(f)$; $i \in I$ } such that there is an algorithm $H_{PQ}$ in **A** such that if a function $f(x)$ that satisfies a predicate $P_i(f)$, then the function $H_{PQ}(f(x))$ satisfies the predicate $Q_i(x)$.

**Theorem 5.3.** There is a number $k$ such that $C_A^Q\{Pr(Q_i(f))\} \leq C_A^P\{Pr(P_i(f))\} + k$ for all $i \in I$.

Let us consider two classes of predicates **P** = { $P_i(f)$; $i \in I$ } and **Q** = { $Q_i(f)$; $i \in I$ } such that there is a Turing machine $T_{PQ}$ that if a function $f(x)$ that satisfies a predicate $P_i(f)$, then the function $T_{PQ}(f(x))$ satisfies the predicate $Q_i(x)$.

**Corollary 5.2.** There is a number $k$ such that $C_T^Q\{Pr(Q_i(f))\} \leq C_T^P\{Pr(P_i(f))\} + k$ for all $i \in I$.

**Proposition 5.2**. If $P(x)$ implies $Q(x)$, then $C\{Pr(Q(f)), T)\} \leq C\{Pr(P(f)), T)\}$ for any Turing machine $T$.

The developed theory of algorithmic problem complexity allows us to develop an inductive hierarchy of problems and to find places in this hierarchy for popular algorithmic problems for Turing machines and other recursive algorithms.

**6. Inductive Turing machines and their hierarchies**

To make this exposition complete, we give a short description of inductive Turing machines. A more detailed exposition is given in [7] or in [10]. The structure of an inductive Turing machine, as an abstract automaton, consists of three components called *hardware*, *software*, and *infware*. Infware is a description and specification of information that is processed by an inductive Turing machine. Computer infware consists of data processed by the computer. Inductive Turing machines are abstract automata working with the same symbolic information in the form of words as conventional Turing machines. Consequently, formal languages with which inductive Turing machines works constitute their infware.

Computer hardware consists of all devices (the processor, system of memory, display, keyboard, etc.) that constitute the computer. In a similar way, an inductive Turing machine *M* has three abstract devices: a *control device A*, which is a finite automaton and controls performance of *M*; a *processor* or *operating device H*, which corresponds to one or several *heads* of a conventional Turing machine; and the *memory E*, which corresponds to the *tape* or tapes of a conventional Turing machine. The memory *E* of the simplest inductive Turing machine consists of three linear tapes, and the operating device consists of three heads, each of which is the same as the head of a Turing machine and works with the corresponding tapes.

The *control device A* is a finite automaton that regulates: the state of the whole machine *M*, the processing of information by *H*, and the storage of information in the memory *E*.

The *memory E* is divided into different but, as a rule, uniform cells. It is structured by a system of relations that organize memory as well-structured system and provide connections or ties between cells. In particular, *input* registers, the *working* memory, and *output* registers of *M* are separated. Connections between cells form an additional structure *K* of *E*. Each cell can contain a symbol from an alphabet of the languages of the machine *M* or it can be empty.

In a general case, cells may be of different types. Different types of cells may be used for storing different kinds of data. For example, binary cells, which have type B, store bits of information represented by symbols 1 and 0. Byte cells (type BT) store information represented by strings of eight binary digits. Symbol cells (type SB) store symbols of the alphabet(s) of the machine *M*. Cells in conventional Turing machines have SB type. Natural number cells, which have type NN, are used in random access machines [1]. Cells in the memory of quantum computers (type QB) store q-bits or quantum bits [23]. Cells of the tape(s) of real-number Turing machines [10] have type RN and store real numbers. When different kinds of devices are combined into one, this new device has several types of memory cells. In addition, different types of cells facilitate modeling the brain neuron structure by inductive Turing machines.

It is possible to realize an arbitrary structured memory of an inductive Turing machine *M*, using only one linear one-sided tape *L*. To do this, the cells of *L* are enumerated in the natural order from the first one to infinity. Then *L* is decomposed into three parts according to the input and output registers and the working memory of *M*. After this, nonlinear connections between cells are installed. When an inductive Turing machine with this memory works, the head/processor is not moving only to the right or to the left cell from a given cell, but uses the installed nonlinear connections.

Such realization of the structured memory allows us to consider an inductive Turing machine with a structured memory as an inductive Turing machine with conventional tapes in which additional connections are established. This approach has many advantages. One of them is that inductive Turing machines with a structured memory can be treated as multitape automata that have additional structure on their tapes. Then it is conceivable to study different ways to construct this structure. In addition, this representation of memory allows us to consider any configuration in the structured memory *E* as a word written on this unstructured tape.

If we look at other devices of the inductive Turing machine *M*, we can see that the processor *H* performs information processing in *M*. However, in comparison to computers, this operational device performs very simple operations. When *H* consists of one unit, it can change a symbol in the cell that is observed by *H*, and go from this cell to another using a connection from **K**. This is exactly what the head of a Turing machine does.

It is possible that the processor *H* consists of several processing units similar to heads of a multihead Turing machine. This allows one to model in a natural way various real and abstract computing systems by inductive Turing machines. Examples of such systems are: multiprocessor computers; Turing machines with several tapes; networks, grids and clusters of computers; cellular automata; neural networks; and systolic arrays.

We know that programs constitute computer software and tell the system what to do (and what not to do). The *software **R*** of the inductive Turing machine *M* also is a program in the form of simple rules:

$$q_h a_i \rightarrow a_j q_k \quad (1)$$
$$q_h a_i \rightarrow c q_k \quad (2)$$
$$q_h a_i \rightarrow a_j q_k c \quad (3)$$

Here $q_h$ and $q_k$ are states of A, $a_i$ and $a_j$ are symbols of the alphabet of M, and c is a type of connection in the memory E.

Each rule directs one step of computation of the inductive Turing machine M. The rule (1) means that if the state of the control device A of M is $q_h$ and the processor H observes in the cell the symbol $a_i$, then the state of A becomes $q_k$ and the processor H writes the symbol $a_j$ in the cell where it is situated. The rule (2) means that the processor H then moves to the next cell by a connection of the type c. The rule (3) is a combination of rules (1) and (2).

Like Turing machines, inductive Turing machines can be deterministic and nondeterministic. For a *deterministic* inductive Turing machine, there is at most one connection of any type from any cell. In a *nondeterministic* inductive Turing machine, several connections of the same type may go from some cells, connecting it with (different) other cells. If there is no connection of the prescribed by an instruction type that goes from the cell that is observed by H, then H stays in the same cell. There may be connections of a cell with itself. Then H also stays in the same cell. It is possible that H observes an empty cell. To represent this situation, we use the symbol $\varepsilon$. Thus, it is possible that some elements $a_i$ and/or $a_j$ in the rules from **R** are equal to $\varepsilon$ in the rules of all types. Such rules describe situations when H observes an empty cell and/or when H simply erases the symbol from some cell, writing nothing in it.

The rules of the type (3) allow an inductive Turing machine to rewrite a symbol in a cell and to make a move in one step. Other rules (1) and (2) separate these operations. Rules of the inductive Turing machine M define the transition function of M and describe changes of A, H, and E. Consequently, they also determine the transition functions of A, H, and E.

A general step of the machine M has the following form. At the beginning of any step, the processor H observes some cell with a symbol $a_i$ (for an empty cell the symbol is $\Lambda$) and the control device A is in some state $q_h$.

Then the control device A (and/or the processor H) chooses from the system **R** of rules a rule r with the left part equal to $q_h a_i$ and performs the operation prescribed by this rule. If there is no rule in **R** with such a left part, the machine M stops functioning. If there are several rules with the same left part, M works as a nondeterministic Turing machine, performing all possible operations. When *A* comes to one of the final states from *F*, the machine *M* also stops functioning. In all other cases, it continues operation without stopping.

For an abstract automaton, as well as for a computer, three things are important: how it receives data, process data and obtains its results. In contrast to Turing machines, inductive Turing machines obtain results even in the case when their operation is not terminated. This results in essential increase of performance abilities of systems of algorithms.

The computational result of the inductive Turing machine *M* is the word that is written in the output register of *M*: when *M* halts while its control device *A* is in some final state from *F*, or when *M* never stops but at some step of computation the content of the output register becomes fixed and does not change although the machine *M* continues to function. In all other cases, *M* gives no result.

**Definition 6.1.** The memory *E* is called *recursive* if all relations that define its structure are recursive.

Here recursive means that there are some Turing machines that decide/build all naming mappings and relations in the structured memory.

**Definition 6.2.** Inductive Turing machines with recursive memory are called *inductive Turing machines of the first order*.

**Definition 6.3.** The memory *E* is called *n-inductive* if all relations that define its structure are constructed by an inductive Turing machine of order *n*.

**Definition 6.4.** Inductive Turing machines with *n*-inductive memory are called *inductive Turing machines of the order n* + 1.

**Definition 6.5.** Two machines are functionally equivalent if they compute the same function.

**Definition 6.6.** An inductive Turing machine *M* is called *finalizing* if it is functionally equivalent to an inductive Turing machine that halts after giving its result.

**Proposition 6.1.** A finalizing inductive Turing machine of the first order is functionally equivalent to a Turing machine.

**Corollary 6.1.** Sequential composition of two finalizing inductive Turing machines of the first order is a finalizing inductive Turing machine of the first order.

However, when the order of machines is larger than one, Proposition 6.1 and its Corollary are not valid in general case.

## 7. Algorithmic Problems for Turing machines and inductive Turing machines

Let us consider popular algorithmic problems for Turing machines considered in popular textbooks, such as [32] or [43], or in fundamental monographs in computer science, such as [44] or [46].

The *Halting Problem* (HP), which is not a purely abstract question because it is equivalent to the similar Halting Problem for computer programs:

given a program $P$ and an input $x$, find if $P$ halts after it starts working with input $x$.

The *Acceptability Problem* (AP): given a Turing machine $T$ and a word $x$, find if $T$ gives a result after it starts working with input $x$.

The *Totality Problem* (TP): given a Turing machine $T$, find if $T$ gives a result for all inputs $x$.

The *Emptiness Problem* (EmP): given a Turing machine $T$, find if $T$ gives no result for all inputs $x$.

The *Language Emptiness Problem* (LEmP): given a Turing machine $T$, find if $T$ the language $L_T$ of $T$ is empty.

The *Equality Problem* (EqP): given Turing machines $Q$ and $T$, find if $Q$ and $T$ define the same function.

The *Language Equality Problem* (LEqP): given Turing machines $Q$ and $T$, find if $L_Q = L_T$.

The *Inclusion Problem* (IcP): given Turing machines $Q$ and $T$, find if $L_Q \subseteq L_T$.

The *Infinity Problem* (IfP): given a Turing machine *T*, find if *T* gives a result for infinite number of inputs *x*.

**Remark 7.1.** In the theory of algorithms, it is proved that these and many other problems related to Turing machines are undecidable by Turing machines.

**Remark 7.2.** It is possible to consider similar problems for other classes of algorithms and automata, for instance, for finite automata or inductive Turing machines.

It is possible consider similar algorithmic problems for inductive Turing machines.

The *Resulting Problem* (RPI): given an inductive Turing machine *M* and a word *x*, find if *M* gives a result after it starts working with input *x*.

The *Totality Problem* (TPI): given an inductive Turing machine *M*, find if *T* gives a result for all inputs *x*.

The *Emptiness Problem* (EmPI): given an inductive Turing machine *M*, find if *M* gives no result for all inputs *x*.

The *Language Emptiness Problem* (LEmPI): given an inductive Turing machine *M*, find if *M* the language $L_M$ of *M* is empty.

The *Equality Problem* (EqPI): given inductive Turing machines *H* and *M*, find if *H* and *M* define the same function.

The *Language Equality Problem* (LEqPI): given inductive Turing machines *H* and *M*, find if $L_H = L_M$.

All these problems have subproblems that are related only to inductive Turing machines of a fixed order. For instance, we have:

The *Resulting Problem* (RPI*n*): given an inductive Turing machine *M* of order *n* and a word *x*, find if *M* gives a result after it starts working with input *x*.

The *Totality Problem* (TPI*n*): given an inductive Turing machine *M* of order *n*, find if *T* gives a result for all inputs *x*.

The *Emptiness Problem* (EmPI*n*): given an inductive Turing machine *M* of order *n*, find if *M* gives no result for all inputs *x*.

Inductive Turing machines of different orders form an infinite hierarchy [7]. We classify problems with respect to this hierarchy.

**Theorem 7.1.** The Resulting Problem (RPI$n$) for inductive Turing machines of order $n$ is undecidable in the class of inductive Turing machines of order $n$.

Proof. To prove this result, we consider only machines with the alphabet $A = \{1, 0\}$. This is not a restriction because it is possible to codify words in any alphabet $C$ by words in $A$ and to simulate an inductive Turing machine that works with words in $C$ by an inductive Turing machine of the same order that works with words in $A$.

In our proof, we use a codification **c**: $IT_n \to A^*$ of inductive Turing machines of order $n$ such that it is possible to reconstruct any inductive Turing machine $T$ of order $n$ by its code **c**($T$). It is possible to find such We also use a pairing function <>: $(A^*) \times (A^*) \to A^*$ that corresponds to each pair $(w, v)$ of words in the alphabet $A$ the word $<w, v>$ in the same alphabet so that different pairs are mapped into different words. It is possible to find how to build pairing functions and codifications of inductive Turing machines, for example, in [10].

We prove the Theorem by contradiction. Namely, assume that there is an inductive Turing machine $D$ of order $n$ that solves the Resulting Problem (RPI$n$) for all inductive Turing machines of order $n$. That is, given a code $<$ **c**($T$), $x >$, the machine $D$ gives 1 as its output when the inductive Turing machine $T$ of order $n$ gives a result being applied to $x$, and gives 0 as its output when the machine $T$ does not give a result being applied to $x$.

Then we build an inductive Turing machine $M$ of order $n$ taking the machine $D$, a simple Turing machine $B$ and a finite automaton $A_C$. Here $B$ is a checking Turing machine such that it checks whether a given word $w$ is equal to **c**($T$) for some inductive Turing machine $T$ of order $n$ and then if this is true it converts $w$ to the word $<$**c**($T$), $w >$. Otherwise, $B$ stops without giving an output. The automaton $A_C$ gives the result 1 for the input 0 and starts an infinite cycle, giving the sequence 1010101 … as its output for all other possible inputs. It is easy to build such inductive Turing machines by standard methods [10].

We build the machine $M$ as the sequential composition $M = B \circ D \circ A_C$. Sequential composition means that the output of each machine in the composition goes as input to each next Turing machine. It is easy to build such composition of machines by standard methods (cf., for example, [43] or [46]). The structure of $M$ is presented in the Figure 1.

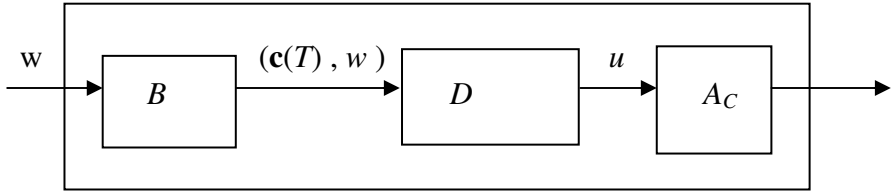

**Figure 1.** The structure of $M = B \circ D \circ A_C$. Here $u$ is some output of $D$.

Now let us find what happens when the inductive Turing machine $M$ receives the word $w = \mathbf{c}(M)$ as its input. This word goes first to the Turing machine $B$, which produces the word $<\mathbf{c}(M), w>$. This pair goes to the Turing machine $D$ as its input. Now we have two options for $M$: $M$ gives a result for the input $w$ or does not give. In the first case, by the definition of inductive Turing machines, the output of $D$ stabilizes on 1 after some moment, which goes to $A_C$ as input. According to its rules, $A_C$ gives the alternating sequence 1010101 … as its output, which means that $M$ does not give a result for $w$ as its input. This contradicts our assumption that $M$ gives a result for the input $w$. So, $M$ does not give a result for $w$ as its input and the output of $D$ stabilizes on 0 after some moment. Thus, 0 starts going to $A_C$ as input. According to its rules, $A_C$ produces 1 as its output each time it receives 0. Consequently, this means that $M$ gives a result for the input $w$. This contradicts our assumption that $M$ does not give a result for $w$ as its input and shows that whatever case we assume for $M$, we come to a contradiction. This contradiction shows that the inductive Turing machine $D$ cannot exist, and thus, Theorem 1 is proved, stating that the Resulting Problem (RPI$n$) for inductive Turing machines of order $n$ is undecidable in the class of inductive Turing machines of order $n$.

**Corollary 7.1.** Complexity with respect to the class of all inductive Turing machines of order $n$ of the Resulting Problem (RPI$n$) for inductive Turing machines of order $n$ is equal to ∞.

**Theorem 7.2.** The Resulting Problem (RPI$n$) for inductive Turing machines of order $n$ is decidable in the class of finalizing inductive Turing machines of order $n + 1$.

<u>Proof</u>. We assume that all words in the alphabet $A$ are enumerated. To build a finalizing inductive Turing machine $M$ of order $n + 1$ that solves the Resulting Problem (RPI$n$) for inductive Turing machines of order $n$, we describe how the structured memory of *M* is organized and how *M* functions. By the definition of inductive Turing machines of order $n + 1$, their structured memory is defined (build) by inductive Turing machines of order *n*.

The structured memory *E* of *M* contains the start cell $c_0$ where the working head is at the beginning of functioning, the finalizing cell $c_1$, the sequence of cells $a_0$, $a_1$, … , $a_n$, … , and cells organized in a standard linear tape *L* of a conventional Turing machine. In addition to connections between cells in *L*, there are two types of connections between other cells: *p* and *t*. Connections of the type *t* connect $c_0$ with $a_0$ and $a_n$ with $a_{n+1}$ for all $n = 1, 2, 3, …$ . Connections of the type *p* are build by a universal inductive Turing machine *U* of order *n*. Namely, *U* connects the cell $a_n$ with $c_1$ if and only if given input *n*, the machine *U* gives a result. Universal inductive Turing machines of order *n* are described in [10]. As connections between cells in *L* and connections of the type *t* are built by finite automata, the memory *E* is n-inductive.

Let us consider an inductive Turing machine *T* of order *n* and a word *w* in the alphabet *A*. Then given the word <*w*, **c**(*T*)> as the input and being in the start state $q_0$, the state of *M* changes to $q_1$ and the working head *h* of the machine *M* comes to the cell $a_n$ where *n* is the number of the word <*w*, **c**(*T*)>. Then the state of *M* changes to $q_2$ and the head *h* writes 0 in the cell $a_n$. After this the rule $q_2 0 \to 0 q_3 c$ is applied if possible. When it is possible to apply this rule, the machine *M* comes to the cell $c_1$, gives 1 as its output and stops. When it is impossible to apply this rule, the machine *M* stops in the cell $a_n$ and gives 0 as its output.

By the definition of the structured memory *E*, the result of *M* is equal to 1 when *T* produces a result, given input *x*, and the result of *M* is equal to 0 when *T* does not produce a result, given input *x*. In such a way, *M* solves the Resulting Problem (RPI*n*) for inductive Turing machines of order *n*. By its construction, the machine *M* is finalizing.

Theorem is proved.

**Remark 7.3.** It is possible to prove Theorem 7.2 using the Hierarchy Theorem for inductive Turing machines from [7]. Another proof may be based on the property that an inductive Turing machine of order $n + 1$ can realize any inductive Turing machine of order *n* as a subprogram.

**Corollary 7.2.** Complexity with respect to the class of all inductive Turing machines of order $n + 1$ of the Resulting Problem (RPI$n$) for inductive Turing machines of order $n$ is equal to a natural number.

### 8. Inductive hierarchies of problems

When in Definitions 4.1, 4.2, and 5.1, **A** is the class of all inductive Turing machines **IT**, then algorithmic complexity with respect to **IT** is called *inductive algorithmic complexity* and problem complexity with respect to **IT** is called *inductive problem complexity*.. We denote all problems that have finite inductive algorithmic complexity by **FIAC**.

When **A** is the class of all inductive Turing machines **IT**$_n$ , then algorithmic complexity with respect to **IT**$_n$ is called *n-inductive algorithmic complexity*. We denote all problems that have finite inductive algorithmic complexity by **FIAC**$_n$.

**Definition 8.1.** a) A problem $P$ has the (*strict*) *inductive order* o($P$) = 0 (so($P$) = 0) if it belongs to the class **FIAC**$_0$, i.e., it can be solved by a Turing machine.

b) A problem $P$ has the (*strict*) *inductive order* o($P$) = $n$ (so($P$) = $n$) if it belongs to the class **FIAC**$_n$ , i.e., it can be solved by an inductive Turing machine of the order $n$ (and does not belong to the class **FIAC**$_{n-1}$ , i.e., it cannot be solved by an inductive Turing machine of the order $n - 1$).

As inductive Turing machines form an infinite hierarchy [7], problems also form an infinite hierarchy called the inductive hierarchy of problems. Theorem 7.2 shows that this is a strict hierarchy.

Properties of the Halting Problem (cf., for example, [10]) and Theorem 7.2 imply the following result.

**Theorem 8.1.** The strict inductive order of the Halting Problem is one, i.e., so(HP) = 1.

Theorems 1 and 2 give the position of the Resulting Problem (RPI$n$) in this hierarchy.

**Theorem 8.2.** so(RPI$n$) = $n + 1$ for all $n$.

Let us consider what orders have algorithmic problems listed in Section 7, i.e., what is the place of these problems in the inductive hierarchy. To do this, we use relations between problems.

**Definition 8.2.** A problem $P$ is *reducible* to a problem $Q$ with respect to a class of algorithms **K** if there is an algorithm $R$ from **K** such that given a solver $D$ to $Q$, algorithm $R$ solves the problem $P$ using results of $D$.

**Remark 8.1.** It is possible that the solver $D$ to $Q$ is simply an oracle [46] or an advice function [2]. Such the solver $D$ can be realized as a hardware in a form of a structured memory.

**Theorem 8.3.** If a problem $P$ is reducible to a problem $Q$ with respect to a class of inductive Turing machine of the order $n$ and the problem $Q$ has the inductive order o($Q$) = $m$, then the problem $P$ has the inductive order o($P$) $\leq m + n$.

Proof is based on a possibility to use the inductive Turing machine of the order $n$ for building the structured memory for an inductive Turing machine that at first, reduces the problem $P$ is reducible to the problem $Q$ and then solves the problem $Q$.

**Remark 8.2.** It is possible that so($P$) < so($Q$) even if $P$ is reducible to $Q$ with respect to a class of Turing machines.

**Definition 8.3.** Problems $P$ and $Q$ are *equivalent* with respect to a class of algorithms **K** if each of them can be reduced to the other by means of algorithms from **K**.

**Lemma 8.1** (cf., for example, [9, 10])**.** The Acceptability Problem is Turing equivalent to the Halting Problem with respect to the class of all Turing machines.

Reductions of different types show that some algorithmic problems have the same order. For instance, Theorems 7.1, 7.2 and 8.3, and Lemma 8.1 imply the following result.

**Corollary 8.1.** The strict inductive order of the Acceptability Problem is one.

The definition of equivalent with respect to the class of all Turing machines algorithmic problems implies the following result.

**Lemma 8.2.** Equivalent with respect to the class of all Turing machines algorithmic problems for Turing machines have the same strict inductive order.

Reducibility of algorithmic problems helps to find their place in the hierarchy.

**Theorem 8.4.** It is possible to reduce the Resulting Problem for inductive Turing machines of the first order to the Totality Problem for Turing machines.

Proof. If $M$ is an inductive Turing machine of the first order and $x$ is a word in the alphabet of $M$, a Turing machine $T_{x,M}$ is corresponded to $M$. We assume that all words in the alphabet of $M$ are ordered and written in a sequence $x_1, x_2, \ldots, x_n, \ldots$ . At the beginning, to find the value $T_{x,M}(x_1)$, the machine $T_{x,M}$ simulates functioning of the machine $M$ working with $x$ as input until $M$ gives two first outputs $M_1(x)$ and $M_2(x)$, and then checks whether $M_2(x)$ is equal to $M_1(x)$. In the case when $M_1(x) \neq M_2(x)$, the machine $T_{x,M}$ gives the result $M_1(x)$ and stops, making $T_{x,M}(x_1)$ equal to $M_1(x)$. Otherwise, $T_{x,M}$ simulates more steps of $M$ until $M$ gives one more output $M_3(x)$, and checks whether $M_3(x)$ is equal to $M_2(x)$. In the case when $M_2(x) \neq M_3(x)$, the machine $T_{x,M}$ gives the result $M_2(x)$ and stops, making $T_{x,M}(x_1)$ equal to $M_2(x)$. This process continues infinitely if and only if all outputs of $M$ coincide with the first output.

To find $T_{x,M}(x_n)$, the machine $T_{x,M}$ simulates functioning of $M$ working with $x$ as input until $M$ gives $n + 1$ outputs $M_1(x), \ldots, M_{n+1}(x)$, and then checks whether $M_{n+1}(x)$ is equal to $M_{n+1}(x)$. In the case when $M_n(x) \neq M_{n+1}(x)$, the machine $T_{x,M}$ gives the result $M_n(x)$ and stops, making the output $T_{x,M}(x_n)$ equal to $M_n(x)$. Otherwise, $T_{x,M}$ simulates more steps of the machine $M$ until $M$ gives one more output $M_{n+2}(x)$, and checks whether $M_{n+2}(x)$ is equal to $M_{n+1}(x)$. In the case when $M_{n+2}(x) \neq M_{n+1}(x)$, the machine $T_{x,M}$ gives the result $M_{n+1}(x)$ and stops, making $T_{x,M}(x_1)$ equal to $M_{n+1}(x)$. This process continues infinitely if and only if all outputs of $M$ coincide with the $n$-th output.

As a result of such definition, the machine $T_{x,M}$ does not compute a total function if and only if $M(x)$ is defined, i.e., $M$ gives the result, working with $x$ as its input.

Consequently, if there is an inductive Turing machine $D$ of the first order that decides whether a given Turing machine computes a total function, then we can build an inductive Turing machine $B$ of the first order that decides whether an arbitrary inductive Turing machine $M$ of the first order applied to $x$ gives a result or not.

Informally, taking a description of M and the word x, the machine B builds $T_{x,M}$ and with a submachine isomorphic to the inductive Turing machine D of the first order that checks whether $T_{x,M}$ computes a total function. Positive answer of the machine D means that being applied to x, the machine M does not give a result and negative answer means that being applied to x, the machine M gives a result.

Theorem is proved.

**Corollary 8.2.** The Totality Problem (TP) for Turing machines is undecidable in the class of all inductive Turing machines of the first order.

**Proposition 8.1.** Inductive Turing machines of the first order can enumerate all Turing machines that do not halt for, at least, one input.

<u>Proof</u>. We build an inductive Turing machine M of the first order that performs such enumeration by building a list of the codes all Turing machines that do not halt, at least, for one input. The place of the code in this list gives the number of the corresponding Turing machine. We assume that all words in the alphabet of M are ordered and written in a sequence $x_1, x_2, \ldots, x_n, \ldots$ . All Turing machines that work with words in the alphabet A are ordered and written in a sequence $T_1, T_2, \ldots, T_n, \ldots$ . The machine M contains following subprograms/submachines: a universal Turing machine U and a machine C that creates codes $< x_m, \mathbf{c}(T_n) >$ according to the algorithm described below. It is possible to find rules for building such machines as C in [32] or [43].

Functioning of M is organized in cycles.

**The first cycle**:

The machine C builds the code $< x_1, \mathbf{c}(T_1) >$, which is given as input to the machine U. The machine U starts simulating the Turing machine $T_1$, which works with input $x_1$. At the same time, M puts $\mathbf{c}(T_1)$ in the list L, which is empty in the beginning of the process. If $T_1$ does not halt after the first step, then M puts $\mathbf{c}(T_2)$ at the end of the list L. If $T_1$ halts after the first step, then M puts $\mathbf{c}(T_2)$ at the beginning of the list L and moves $\mathbf{c}(T_1)$ to the end of the list, i.e., puts $\mathbf{c}(T_1)$ after $\mathbf{c}(T_2)$. Then the machine M goes to the second cycle.

**The second cycle**:

If $T_1$ does not halt in the first cycle, the machine $C$ builds codes $< x_2 , \mathbf{c}(T_1) >, < x_1 , \mathbf{c}(T_2) >$, and $< x_2 , \mathbf{c}(T_2) >$, gives these codes as inputs to the machine $U$, and $U$ starts simulating both $T_1$ and $T_2$ with inputs $x_1$ and $x_2$, making two steps in each case.

In the case, when $T_1$ halts in the first cycle, the machine $C$ builds codes $< x_2 , \mathbf{c}(T_1) >, < x_1 , \mathbf{c}(T_2) >$, and $< x_2 , \mathbf{c}(T_2) >$, gives these codes as inputs to the machine $U$, and $U$ starts simulating $T_1$ with inputs $x_2$ and $T_2$ with inputs $x_1$ and $x_2$, making two steps in each case.

If both $T_1$ and $T_2$ do not do not halt in these simulations, $M$ does not change order in the list $L$ and adds to it the code $\mathbf{c}(T_3)$ at the end.

If $T_1$ halts and $T_2$ does not halt in these simulations, then $M$ puts $\mathbf{c}(T_2)$ at the beginning of the list $L$ and moves $\mathbf{c}(T_1)$ to the end of the list, i.e., puts $\mathbf{c}(T_1)$ after $\mathbf{c}(T_2)$. In addition, the machine $M$ inserts the code $\mathbf{c}(T_4)$ between $\mathbf{c}(T_2)$ and $\mathbf{c}(T_1)$.

If $T_2$ halts and $T_1$ does not halt in these simulations, then $M$ puts $\mathbf{c}(T_1)$ at the beginning of the list $L$ and moves $\mathbf{c}(T_2)$ to the end of the list, i.e., puts $\mathbf{c}(T_2)$ after $\mathbf{c}(T_1)$. In addition, the machine $M$ inserts the code $\mathbf{c}(T_4)$ between $\mathbf{c}(T_1)$ and $\mathbf{c}(T_2)$.

If both $T_1$ and $T_2$ halt in these simulations, then $M$ puts $\mathbf{c}(T_3)$ at the beginning of the list $L$ and moves $\mathbf{c}(T_1)$ and $\mathbf{c}(T_2)$ to the end of the list, i.e., puts $\mathbf{c}(T_1)$ and $\mathbf{c}(T_2)$ after $\mathbf{c}(T_3)$.

Then the machine $M$ goes to the third cycle.

**The third cycle**:

If both $T_1$ and $T_2$ do not halt in the second cycle, then the machine $C$ builds codes $< x_3 , \mathbf{c}(T_1) >, < x_3 , \mathbf{c}(T_2) >, < x_1 , \mathbf{c}(T_3) >, < x_2 , \mathbf{c}(T_3) >$, and $< x_3 , \mathbf{c}(T_3) >$, gives these codes as inputs to the machine $U$, and $U$ starts simulating $T_1$, $T_2$ and $T_3$ with inputs $x_1$, $x_2$ and $x_3$, making three steps in each case.

In case when $T_1$ and/or $T_2$ halt in the second cycle, $U$ simulates only those of machines $T_1$ and $T_2$ with such inputs $x_1$ and $x_2$ for which the corresponding machine did not stop. For instance, if $T_1$ stops for input $x_1$, then $U$ stops simulating $T_1$ with this input for all consequent cycles. The machine $U$ also simulates machines $T_1$ and $T_2$ with input $x_3$ and $T_3$ with inputs $x_1$, $x_2$, and $x_3$, making three steps in each case.

After performing all three steps of simulation for all given machines and inputs, the machine $M$ starts working with the list $L$. If neither of the machines halts in these simulations, then the machine *M* does not change order in the list *L* and adds to it the code **c**($T_4$) at the end.

Otherwise, codes of those of the machines $T_1$, $T_2$ and $T_3$ that halt in these simulations are moved to the end of the list, while the code **c**($T_4$) is inserted before those codes that are moved.

When all three Turing machines $T_1$, $T_2$, and $T_3$ halt in these simulations, the machine *M* puts **c**($T_4$) at the beginning of the list *L* and moves **c**($T_1$), **c**($T_2$) and **c**($T_3$) to the end of the list, i.e., puts **c**($T_1$), **c**($T_2$) and **c**($T_3$) after **c**($T_4$).

Then the machine *M* goes to the next cycle.

**The $n^{th}$ cycle**:

The machine *C* simulates Turing machines $T_1$, $T_2$, $T_3$, ... , $T_n$ with those of the inputs $x_1$, $x_2$, $x_3$, ... , $x_n$ for the corresponding Turing machine $T_k$ did not stop in one of the previous cycles, i.e., if the machine $T_k$ halts after *m* steps on input $x_i$ and $m < n$, then the machine $T_k$ is not simulated with input $x_i$ in this cycle. The machine *C* simulates machines $T_1$, ... , $T_n$ with relevant inputs, making *n* steps in each case.

After performing all *n* steps of simulation for all given machines and inputs, the machine *M* starts working with the list *L*. At first, the machine *M* adds the code **c**($T_{n+1}$) to the end of the list. If neither of the machines halts in these simulations, then the machine *M* does not change order in the list *L*. Otherwise, codes of those of the machines $T_k$ that halt on all their inputs in these simulations are moved to the end of the list in the same order as they were in the list.

Then the machine *M* goes to the next cycle.

Given a number *k* as input to the machine *M*, this machine starts the whole simulation process described above. When the first code **c**($T_r$) appears at the place *k* of the list *L*, the machine *M* gives this code as its output and repeats doing this in each cycle.

The procedure is such that if a Turing machine $T_r$ computes a total function, the code **c**($T_r$) will be always moved to the end of the list *L* and in some cycle it will forever disappear from the output of *M* for any number *k*. At the same time, if a machine $T_q$ that does not stop for some input, then after several cycles (may be, even after one cycle), the code $c(T_r)$ stops moving in the list $L$ and consequently, the output of $M$ stabilizes after some cycle for any input.

Rules of functioning for the machine $M$ are finite and constructive. So, it is possible to build an inductive Turing machine $M$ that satisfies these rules. As it is demonstrated, for each input the machine $M$ computes the code of some Turing machine $T_r$ that does not stop for some input and codes for all such Turing machines will be computed.

Proposition is proved.

**Theorem 8.5.** The Totality Problem (TP) for Turing machines is decidable in the class of inductive Turing machines of the second order.

<u>Proof</u>. We assume that all Turing machines, or more exactly, their codes, are effectively enumerated. Thus, the number of a Turing machine is unique and allows one to reconstruct this machine. To build an inductive Turing machine $H$ of the second order that solves (TP), we describe how the structured memory of $H$ is organized and how $H$ functions. By the definition of inductive Turing machines, the structured memory is build by the inductive Turing machine $M$ of the first order in the following way. The memory of $H$ contains cells $c_1, c_2, \ldots, c_n, \ldots$ . Each of these cells can contain one symbol from the alphabet $A$. The machine $M$ combines these cells into sets (hypercells) $a_1, a_2, \ldots, a_n, \ldots$ so that $a_j$ contains the number $i$ of the Turing machine $T_i$ the code $c(T_i)$ of which is at the place $j$ in the list $L$ constructed by the inductive machine $M$ in Proposition 8.1.

The machine $H$ works with numbers of Turing machines. Given a number $m$, the head of the machine $H$ starts going from the hypercell $a_1$ to $a_2$ to $a_3$ and so on. Coming to a hypercell $a_n$, the head checks whether $a_n$ contains the code $c(T_m)$ of the Turing machine $T_m$. If $a_n$ does not contain the code $c(T_m)$, the head goes to the next hypercell $a_{n+1}$, while the machine $H$ gives 1 as its output. When the head finds the code $c(T_m)$, the machine $H$ gives 0 as its output and stops.

The list $L$ constructed by the inductive machine $M$ in Proposition 8.1 contains codes of all Turing machines that compute non-total functions. Consequently, the machine $H$ solves the Totality Problem (TP) for Turing machines, giving 0 when the tested Turing machine $T_m$ computes a non-total function and 0 when the tested Turing machine $T_m$ computes a total function.

Theorem is proved.

This result allows us to find the place of the Totality Problem for Turing machines in the inductive hierarchy.

**Theorem 8.6.** The strict inductive order so(TP) of the Totality Problem (TP) for Turing machines is two.

Indeed, by Corollary 8.2, The Totality Problem for Turing machines is undecidable in the class of all inductive Turing machines of the first order. At the same time, by Theorem 8.5, (TP) is decidable in the class of all inductive Turing machines of the second order. Consequently, so(TP) = 2.

**Proposition 8.2.** Problems (IfP) and (TP) are equivalent with respect to the class of all Turing machines.

<u>Proof</u>. a) Let $T$ be a Turing machine and **K** be some class of algorithms. By Theorem V from ([46] Ch.5), it is possible to construct a Turing machine $N_T$ that enumerates all elements from the range R($T$) of $T$. As it is possible to assume that all words in the alphabet $A$ of $T$ are also enumerated a Turing machine, we may presuppose that the Turing machine $N_T$ enumerates all elements from the range R($T$) of $T$ by words in the alphabet $A$ of $T$. Naturally, this machine defines a total function if and only if R($T$) is infinite. Thus, if the problem (TP) is solvable in **K**, then the problem (IfP) also is solvable in **K**.

b) Let $T$ be a Turing machine and all words in the alphabet $A$ are organized in the sequence $x_1$, $x_2$, ... , $x_n$, ... . We define $M_T$ as a Turing machine that given $x = x_n$ as its input computes all values $T(x_1)$, $T(x_2)$, ... , $T(x_{n-1})$, $T(x_n)$ when it is possible and then stops, giving $T(x_n)$ as its output. When $T$ does not give a result for $x_i$ as its input for some $i < n + 1$, then $M_T$ also does not give a result for $x_n$ as its input. Thus, the machine $M_T$ has infinite range if and only if $T$ computes a total function. Consequently, if the problem (IfP) is solvable in **K**, then the problem (TP) also is solvable in **K**.

Proposition is proved.

Proposition 8.2 and Lemma 8.2 imply the following theorem.

**Theorem 8.7.** The strict inductive order so(IfP) of the Infinity Problem for Turing machines is two.

In addition, we have the following result.

**Theorem 8.8.** The strict inductive order so(EmP) of the Emptiness Problem (EmP) is one.

Proof. At first, we build an inductive Turing machine $K$ of the first order that solves the Emptiness Problem for Turing machines. This machine works in the following way.

As before, we assume that all words in the given alphabet $A$ are ordered and written in a sequence $x_1$, $x_2$, … , $x_n$, … . In addition, the machine $K$ contains a copy of a universal Turing machine and thus, is able to simulate any Turing machine $T$. Given the code $\mathbf{c}(T)$ of some Turing machine $T$, the machine $K$ starts simulating $T$ with inputs $x_1$, $x_2$, … , $x_n$, … . Simulation goes in cycles. In the cycle number $n$, the machine $K$ simulates $n$ steps of functioning of $T$ with inputs $x_1$, $x_2$, … , $x_n$ if $K$ has not stop in any of the previous $(n-1)$ cycles because $T$ gives some result in one of those cycles. When in some cycle, simulation of $T$ halts, giving some result, the machine $K$ gives 0 as its output and stops. When $K$ completes a cycle without halting, gives 1 as its output and goes to the next cycle. In such a way, the machine $K$ solves the Emptiness Problem for an arbitrary Turing machine $T$.

At the same time, from the theory of Turing machines, we know (cf., [32] or [46]) that the Emptiness Problem is undecidable in the class of all Turing machines. Consequently, so(TP) = 1.

Theorem is proved.

The Emptiness Problem (EmP) is Turing equivalent to the Language Emptiness Problem (LEmP). This implies the following result.

**Corollary 8.3.** The strict inductive order so(LEmP) of the Language Emptiness Problem is one.

## 9. Conclusion

In this work, problems are classified in order to build measures of problem complexities. They are built as dual static complexity measures, at first, with respect to Turing machines and then with respect to arbitrary classes of algorithms. Optimal problem complexities are found. Different properties of problem complexities are obtained. In addition, superrecursive classes of problem complexities are separated. Complexities of some well-known problems, such as the Halting Problem for Turing machines, are determined in the context of inductive hierarchy of problems.

It is possible also to introduce and study hard and complete problems for these classes.

According to a general approach (cf. [10, 32]), a problem *p* is *hard* for a class K of algorithms if any problem decidable in K can be reduced to the problem *p* by algorithms from K.

In a similar way, a problem *p* decidable in a class K is *complete* for K if it is hard for the class K.

In other words, complete problems for a class are hard problems that belong to the same class.

In particular, it would be appealing to find if the Halting Problem is complete for the class of all inductive Turing machines of the first order.

The developed theory of algorithmic problem complexity is closely related to problems of information measurement. For instance, it is proved (cf., for example, [40]) that Kolmogorov complexity allows one to derive a good approximation to the classical Shannon's measure of information. Principles of the general theory of information [6] show that problem complexity in a general case might be useful for building different measures of information and for finding useful properties and regularities of information and information processes.

The approach to complexity in this work is semi-axiomatic because it is axiomatic with respect to classes of algorithms, but works only for a given complexity measure – the length of a word/program/algorithm. It would be important and interesting to study problem complexity in an axiomatic form where both complexity measures and classes of algorithms are defined by some axioms (properties). In a similar way as it is done in [10] for algorithmic complexity of words, axiomatic problem complexity would extend the scope of applications.

Another direction for future research is to study dynamic problem complexity. Examples of dynamic problem complexities are the least, in some sense, time or space needed to solve a given kind of problems.